\documentclass[12pt]{article}
\usepackage{amssymb, amsmath}
\usepackage{amsfonts}
\usepackage{color}
\newcommand{\Section}[1]%
{\section{#1}\setcounter{equation}{0}%
\setcounter{theorem}{0}}
\newtheorem{theorem}{Theorem}

\newtheorem{pro}[theorem]{Proposition}
\def\re{\mathbb{R}}
\def\co{\mathbb{C}}

\def\ze{\mathbb{Z}}

\newenvironment{proof}[1]%
{\par\noindent{\em #1:\ }}%
{~\rule{2mm}{2mm}\par\bigskip}
\begin{document}
%%%%%%%%%%%%%%%%%%%%%%%%%%%%%%%%%%%%%%%%%%%
%%%%%%%%%%%%%%%%%%%%%%%%%%%%%%%%%%%%%%%%%%%
%%%%%%%%%%%%%%%%%%%%%%%%%%%%%%%%%%%%%%%%%%%
\newpage\thispagestyle{empty}
{\topskip 2cm
\begin{center}
{\Large\bf Dispersion Relations of Nambu-Goldstone Modes\\} 
\bigskip\bigskip
{\Large Tohru Koma
\footnote{\small \it Department of Physics, Gakushuin University (retired), Mejiro, Toshima-ku, Tokyo 171-8588, JAPAN}
%{\small\tt e-mail: tohru.koma@gakushuin.ac.jp}
\\}
\end{center}
\vfil
\noindent
{\bf Abstract:} We study the energy-momentum relations of the Nambu-Goldstone modes in 
quantum antiferromagnetic Heisenberg models on a hypercubic lattice. 
This work is a sequel to the previous series about the models. 
We prove that the Nambu-Goldstone modes show a linear dispersion relation for the small momenta, 
i.e., the low energies of the Nambu-Goldstone excitations are proportional to the momentum of the spin wave 
above the infinite-volume ground states with symmetry breaking. 
Our method relies on the upper bounds for the susceptibilities which are derived from 
the reflection positivity of the quantum Heisenberg antiferromagnets. 
The method is also applicable to other systems when the corresponding upper bounds 
for the susceptibilities are justified.    
\par
%%%%%%%%%%%%%%%%%%%%%%%%%%%%%%%%%%%%%%%%%%%%%%%%%%%%%%%%%%%%%%%%%%%%%%%%%%%
%\noindent
%\bigskip
%\hrule
%\bigskip
%%%%%%%%%%%%%%%%%%%%%%%%%%%%%%%%%%%%%%%%%%%%%%%%%%%%%%
\vfil}
%\newpage
%%%%%%%%%%%%%%%%%%%%%%%%%%%%%%%%%%%%%%%%%%%

%%%%%%%%%%%%%%%%%%%%%%%%%%%%%%%%%%%%%%%%%%%%%%%%%%%%%%
\Section{Introduction}
\label{Intro}

In contrast to the quantum ferromagnetic Heisenberg models, 
the ground state of the quantum antiferromagnets is mysterious. 
In fact, although the ground state of the quantum antiferromagnets for a finite volume is unique \cite{Marshall,LM}, 
an infinite-volume ground state of the same model exhibits a spontaneous staggered magnetization \cite{KomaTasaki}.  
As is well known, the classical N\'eel state exhibits a staggered magnetization 
but is not a ground state of the quantum antiferromagnets. In other words, the unique finite-volume ground state retains 
the symmetry of the model for any finite volume, but the infinite-volume ground state breaks the symmetry. 
This phenomenon can be explained by the quasi-degeneracy of the finite-volume ground state \cite{KomaTasaki2,Tasaki}. 
Namely, there appear many low-lying eigenstates whose excitation energy is 
very close to the energy of the symmetric ground state of the finite-volume Hamiltonian, 
and these low-lying eigenstates yield a set of symmetry-breaking ground states in the infinite-volume limit 
by forming linear combinations of the low-lying eigenstates and the symmetric ground state. 

This idea of the quasi-degeneracy explains well the emergence of the infinite-volume ground states 
with symmetry breaking. However, there arises another question as follows: 
How do we obtain the low-energy excitations above an infinite-volume ground state?  
In other words, how do we distinguish the low-energy excitations from the low-lying eigenstates 
which yield a set of the infinite-volume ground states? 
Clearly, the low energy limit of a low-energy excitation above an infinite-volume ground state 
is degenerate to the infinite-volume ground state.  
The excitation spectrum for low energy states above the ground state has been often computed 
by using trial wavefunctions within Bijl-Feynman single-mode approximation \cite{Bijl,Feynman,Takada,Stringari,Momoi}.
Such a trial state is not guaranteed to be orthogonal to the sector of the infinite-volume ground states.   

In order to avoid this difficulty, we can use a spectral projection which is defined by the time evolution of 
local operators \cite{AL,KomaS} for the purpose of obtaining the spectrum of the low-energy excitations above the infinite-volume 
ground state with symmetry breaking. By the spectral projection, the excitation energies of a trial state 
are restricted to strictly positive energies in the infinite-volume limit. 

In this paper, we study the energy-momentum relations of the Nambu-Goldstone modes \cite{Nambu,NJL,Goldstone,GSW} in 
quantum antiferromagnetic Heisenberg models on a $d$-dimensional hypercubic lattice with $d\ge 2$. 
As a result, we prove that the Nambu-Goldstone modes show a linear dispersion relation for the small momenta, 
i.e., the low energies of the Nambu-Goldstone excitations are proportional to the momentum of the spin wave 
above the infinite-volume ground states with symmetry breaking. 
Our method relies on the upper bounds for the susceptibilities which are derived from 
the reflection positivity \cite{DLS,JNFP,KLS} of the quantum Heisenberg antiferromagnets. 
The method is also applicable to other systems when the corresponding upper bounds 
for the susceptibilities are justified. 
The present work is a sequel to the previous series \cite{KomaTasaki,KomaTasaki2,KomaM,KomaP} 
about the quantum Heisenberg antiferromagnets. 

The present paper is organized as follows: In the next Section~\ref{Sec:ModelResult}, the precise definition of 
the quantum antiferromagnetic Heisenberg models and the main Theorem~\ref{mainTheorem} are given. 
Section~\ref{Sec:Preliminary} is devoted to a preliminary for giving the proof of the main Theorem~\ref{mainTheorem}. 
In Sec.~\ref{Sec:Excitations}, we recall the standard thermodynamic formalism \cite{AL,KomaS} 
to deal with excitation energies above infinite-volume ground states. 
If a reader is familiar with the formalism or is not interested in the mathematical justification, 
then one can skip Sec.~\ref{Sec:Excitations}. The proof of the main Theorem~\ref{mainTheorem} is given in Sec.~\ref{Sec:proof}. 
In Sec.~\ref{Sec:DifferentNGmode}, we present a different type Nambu-Goldstone mode from that in the main Theorem~\ref{mainTheorem}.  
The set of these two types of the excitations is an analogue of the des Cloizeaux-Pearson mode \cite{dCPmode} in  
the one-dimensional quantum Heisenberg antiferromagnet. 
Appendices~\ref{Sec:QuasiLocal} and \ref{Sec:continuity} are devoted to technical estimates.

%%%%%%%%%%%%%%%%%%%%%%%
\Section{Model and result}
\label{Sec:ModelResult}

We consider quantum Heisenberg antiferromagnets which have reflection positivity \cite{DLS,KLS}. 
The extension of our method to anisotropic antiferromagnets is relatively straightforward \cite{KLS2,KuboKishi}. 
More precisely, we can treat the Hamiltonian $H_{0,B}^{(\Lambda)}$ of (\ref{ham}) below with an additional Ising term. 

Let $\Gamma$ be a finite subset of the $d$-dimensional hypercubic lattice $\ze^d$, i.e., $\Gamma\subset\mathbb{Z}^d$,  
with $d\ge 1$. For each site $x=(x^{(1)},x^{(2)},\ldots,x^{(d)})\in\Gamma$, 
we associate three component quantum spin operator ${\bf S}_x=(S_x^{(1)},S_x^{(2)},S_x^{(3)})$ 
with magnitude of spin, $S=1/2,1,3/2,2,\ldots$. More precisely, the spin operators, $S_x^{(1)}, S_x^{(2)}, S_x^{(3)}$, 
are $(2S+1)\times(2S+1)$ matrices at the site $x$. They satisfy the commutation relations, 
$$
[S_x^{(1)},S_x^{(2)}]=iS_x^{(3)}, \quad [S_x^{(2)},S_x^{(3)}]=iS_x^{(1)}, 
\quad \mbox{and} \quad [S_x^{(3)},S_x^{(1)}]=iS_x^{(2)},
$$
and $(S_x^{(1)})^2+(S_x^{(2)})^2+(S_x^{(3)})^2=S(S+1)$ for $x\in\Gamma$. 
For the finite lattice $\Gamma$, the whole Hilbert space is given by 
$$
\mathfrak{H}_\Gamma=\bigotimes_{x\in\Gamma} \co^{2S+1}.
$$
More generally, the algebra of observables on $\mathfrak{H}_\Gamma$ is given by 
$$
\mathfrak{A}_\Gamma:=\bigotimes_{x\in\Gamma}M_{2S+1}(\co),
$$
where $M_{2S+1}(\co)$ is the algebra of $(2S+1)\times(2S+1)$ complex matrices. 
When two finite lattices, $\Gamma_1$ and $\Gamma_2$, satisfy $\Gamma_1\subset\Gamma_2$, 
the algebra $\mathfrak{A}_{\Gamma_1}$ is embedded in $\mathfrak{A}_{\Gamma_2}$ by 
the tensor product $\mathfrak{A}_{\Gamma_1}\otimes I_{\Gamma_2\backslash\Gamma_1}\subset 
\mathfrak{A}_{\Gamma_2}$ with the identity $I_{\Gamma_2\backslash\Gamma_1}$. 
The local algebra is given by 
$$
\mathfrak{A}_{\rm loc}=\bigcup_{\Gamma\subset\ze^d:|\Gamma|<\infty}\mathfrak{A}_{\Gamma},
$$
where $|\Gamma|$ is the number of the sites in the finite lattice $\Gamma$. 
The quasi-local algebra is defined by the completion of the local algebra $\mathfrak{A}_{\rm loc}$ 
in the sense of the operator-norm topology. 

Consider a $d$-dimensional finite hypercubic lattice, 
\begin{equation}
\label{Lambda}
\Lambda:=\{-L+1,-L+2,\ldots,-1,0,1,\ldots,L-1,L\}^d\subset\mathbb{Z}^d,
\end{equation}
with a large positive integer $L$ and $d\ge 2$. 
The Hamiltonian of the quantum antiferromagnetic Heisenberg model on the lattice $\Lambda$ is given by 
\begin{equation}
\label{ham}
H_{0,B}^{(\Lambda)}:=\sum_{x,y\in\Lambda:|x-y|=1}[S_x^{(1)}S_y^{(1)}+S_x^{(2)}S_y^{(2)}+S_x^{(3)}S_y^{(3)}]
-B\sum_{x\in\Lambda} (-1)^{x^{(1)}+x^{(2)}+\cdots+x^{(d)}}S_x^{(1)},
\end{equation}
where we impose the periodic boundary condition so that the system has the translational invariance in 
all the directions with the period $2$, and $B\ge 0$. Further, in the following, we assume $d\ge 2$ 
because the phase transitions do not occur in the one-dimensional system \cite{Shastry,TTI}. 

To begin with, we show that the ground state of the Hamiltonian $H_{0,B}^{(\Lambda)}$ is unique for $B>0$. 
Following \cite{Tanaka}, we introduce a unitary transformation, 
\begin{equation}
U^{(\Lambda)}:=\prod_{x\; :\; x^{(1)}+x^{(2)}+\cdots+x^{(d)}=\mbox{\footnotesize odd}}\exp[i\pi S_x^{(3)}]. 
\end{equation}
One obtains 
\begin{eqnarray}
\label{tildeham}
\tilde{H}_{0,B}^{(\Lambda)}&:=&[U^{(\Lambda)}]^\ast H_{0,B}^{(\Lambda)}U^{(\Lambda)}\nonumber\\
&=&\sum_{x,y\in\Lambda:|x-y|=1}\left\{-\frac{1}{2}[S_x^{(+)}S_y^{(-)}+S_x^{(-)}S_y^{(+)}]+S_x^{(3)}S_y^{(3)}]\right\}
-\frac{B}{2}\sum_{x\in\Lambda} [S_x^{(+)}+S_x^{(-)}],\nonumber\\
\end{eqnarray}
where $S_x^{(\pm)}:=S_x^{(1)}\pm iS_x^{(2)}$. Clearly, this Hamiltonian has the translational invariance with the period $1$ 
in contrast to $H_{0,B}^{(\Lambda)}$.  
Therefore, the ground state $\tilde{\Phi}_{0,B}^{(\Lambda)}$ of 
this Hamiltonian $\tilde{H}_{0,B}^{(\Lambda)}$ is unique with the momentum zero for $B>0$ 
by the Perron-Frobenius theorem (See, e.g., Appendix~A.4.1 in the book \cite{Tasakibook}). 
We write $\Phi_{0,B}^{(\Lambda)}$ for the ground-state vector of the Hamiltonian $H_{0,B}^{(\Lambda)}$ of (\ref{ham})
with the normalization $\Vert\Phi_{0,B}^{(\Lambda)}\Vert=1$. Then, one has 
\begin{equation}
\Phi_{0,B}^{(\Lambda)}=U^{(\Lambda)}\tilde{\Phi}_{0,B}^{(\Lambda)}.
\end{equation} 
The infinite-volume ground state is given by 
\begin{equation}
\label{omega0}
\omega_0(\cdots):=\mbox{weak}^\ast\mbox{-}\lim_{B\searrow 0}\mbox{weak}^\ast\mbox{-}\lim_{\Lambda\nearrow\ze^d}
\left\langle\Phi_{0,B}^{(\Lambda)},(\cdots)\Phi_{0,B}^{(\Lambda)}\right\rangle  
\end{equation} 
with the infinitesimally weak field $B$ for the symmetry breaking. 

Consider an observable $a^{(n)}$ which is given by 
\begin{equation}
\label{a0}
a^{(n)}=\sum_{x} f^{(n)}(x) S_x^{(2)},
\end{equation}
where the complex-valued function $f^{(n)}(x)$ has a compact support, and satisfies 
$f^{(n)}(x)\rightarrow f(x)$ as $n\rightarrow\infty$ for a function,  
\begin{equation}
\label{f(x)}
f(x)=\frac{1}{(2\pi)^d}\int_{-\pi}^{+\pi}dk^{(1)} \int_{-\pi}^{+\pi}dk^{(2)}\cdots \int_{-\pi}^{+\pi}dk^{(d)} \hat{f}(k)e^{ikx}, 
\end{equation}
with a complex-valued function $\hat{f}(k)$ of the momentum $k=(k^{(1)},k^{(2)},\ldots,k^{(d)})$. 
Here, $kx$ denotes the inner product $kx=k^{(1)}x^{(1)}+k^{(2)}x^{(2)}+\cdots+k^{(d)}x^{(d)}$. 
We take the function $\hat{f}(k)$ to be infinitely differentiable so that the function $f(x)$ in the infinite-volume limit 
rapidly decays with distance from the origin $x=0$. More specifically, we choose the function $\hat{f}$ such that 
\begin{equation}
\label{supphatf}
{\rm supp}\; \hat{f}\subset \{k\;|\; {|p|}/{2}\le |k|\le |p|\}
\end{equation}
with a momentum $p$ satisfying $0<|p|<\kappa$ with a small $\kappa>0$. Namely, the wavepacket has 
the small momenta whose strengths are all nearly equal to $|p|$.    

Next, in order to introduce an energy cutoff, we consider the time evolution \cite{BR} of local 
operator $a^{(n)}$, 
$$ 
\tau_{t,B}^{(\Lambda)}(a^{(n)}):=\exp[iH_{0,B}^{(\Lambda)}t]a^{(n)}\exp[-iH_{0,B}^{(\Lambda)}t], 
$$
and 
$$
\tau_{\ast g,B}^{(\Lambda)}(a^{(n)}):=\int_{-\infty}^{+\infty}dt\; g(t)\tau_{t,B}^{(\Lambda)}(a^{(n)}),
$$
where the function $g$ is the Fourier transform of an infinitely differentiable function $\hat{g}$ 
with a compact support, i.e., $\hat{g}\in C_0^\infty(\re)$. 
We choose the real-valued function $\hat{g}$ 
to satisfy ${\rm supp}\;\hat{g}\subseteq(\epsilon,\gamma)$ with two positive parameters, $\epsilon$ and $\gamma$,  
which satisfy $0<\epsilon<\gamma$. Write \cite{BR}
$$
\tau_{\ast g,0}(a):=\lim_{n\nearrow\infty}\lim_{B\searrow 0}\lim_{\Lambda\nearrow\ze^d}\tau_{\ast g,B}^{(\Lambda)}(a^{(n)}),
$$
where $a=\lim_{n\nearrow\infty}a^{(n)}$ and the sequences in the double limit about $\Lambda$ and $B$ 
are taken to be the same as those in the ground state $\omega_0$ of (\ref{omega0}). 
However, this converges to the desired value $\tau_{\ast g,0}(a)$, irrespective of the subsequences \cite{NOS}. 
In passing, we note the following: The observable $a$ is given by 
\begin{equation}
a=\sum_{x\in\ze^d} f(x) S_x^{(2)},
\end{equation}
{from} (\ref{a0}). Since the function $f(x)$ of (\ref{f(x)}) has the momentum distribution (\ref{supphatf}), 
the observable $a$ is interpreted as a wavepacket operator of the spin wave with small momenta 
whose strengths are nearly equal to $|p|$.  

Our aim is to estimate the energy of the excitation above the infinite-volume ground state which is given by 
\begin{equation}
\label{DeltaE}
\Delta E:=\lim_{\Lambda\nearrow\ze^d}\frac{\omega_0([\tau_{\ast g,0}(a)]^\ast[H_{0,0}^{(\Lambda)},\tau_{\ast g,0}(a)])}
{\omega_0([\tau_{\ast g,0}(a)]^\ast\tau_{\ast g,0}(a))}\ge \epsilon,
\end{equation}
where $H_{0,0}^{(\Lambda)}$ is given by $H_{0,B}^{(\Lambda)}$ with $B=0$. 
The lower bound in the right-hand side is obtained from the definition of the function $\hat{g}$ 
which gives a projection onto the subspace with the energy interval $[\epsilon,\gamma]$. 
More specifically, we choose the real-valued function $\hat{g}\in C_0^\infty(\re)$ such that it satisfies the following conditions: 
\begin{equation}
0\le \hat{g}(E) \le 1 \ \mbox{for} \ \mbox{all}\ E\in\re, \quad \hat{g}(E)=1 \ \mbox{for} \ 2\epsilon \le E \le \gamma -\Delta\gamma, 
\end{equation} 
and
\begin{equation}
\hat{g}(E)=0 \ \mbox{for} \ E\le \epsilon \ \mbox{or} \ E\ge \gamma 
\end{equation}
with a constant $\Delta\gamma>0$ satisfying $2\epsilon<\gamma-\Delta\gamma$. 

The staggered  magnetization is given by 
\begin{equation}
\label{mBLambda}
m_B^{(\Lambda)}:=\frac{1}{|\Lambda|}\sum_{x\in\Lambda} (-1)^{x^{(1)}+x^{(2)}+\cdots+x^{(d)}}
\langle \Phi_{0,B}^{(\Lambda)},S_x^{(1)}\Phi_{0,B}^{(\Lambda)}\rangle.
\end{equation}
The spontaneous magnetization $m_{\rm s}$ in the infinite-volume limit is given by 
\begin{equation}
m_{\rm s}:=\lim_{B\searrow 0}\lim_{\Lambda\nearrow\ze^d} m_B^{(\Lambda)}.
\end{equation}
We write 
$$
\omega_{\rm wp}(\cdots):=\frac{\omega_0([\tau_{\ast g,0}(a)]^\ast(\cdots)\tau_{\ast g,0}(a))}
{\omega_0([\tau_{\ast g,0}(a)]^\ast\tau_{\ast g,0}(a))}
$$
for short.  

\begin{theorem}
\label{mainTheorem}
Suppose that the spontaneous magnetization $m_{\rm s}$ is non-vanishing, i.e., $m_{\rm s}>0$. 
Then, there exist positive constants, $v_{\rm min}$ and $v_{\rm max}$, such that  
the excitation energy $\Delta E$ of the wavepacket $\omega_{\rm wp}$ above the infinite-volume ground state $\omega_0$ satisfies  
\begin{equation}
v_{\rm min}|p|\le \Delta E\le v_{\rm max}|p|
\end{equation}
with the strength $|p|$ of the momentum $p$ which is the approximate momentum of the wavepacket with a compact, 
narrow distribution (\ref{supphatf}) of $\hat{f}$ in the momentum space. 
\end{theorem}

\medskip
\noindent
{\it Remark:} (i) The excitation in Theorem~\ref{mainTheorem} has the momenta around the zero momentum $p=0$.  
We can also construct a low energy excitation around the momentum $Q=(\pi,\pi,\ldots,\pi)$,  
as will show in Sec.~\ref{Sec:DifferentNGmode}. 
The set of these two types of the excitations is an analogue of the des Cloizeaux-Pearson mode \cite{dCPmode} in  
the one-dimensional quantum Heisenberg antiferromagnet. 
\smallskip

\noindent
(ii) Our method relies on the upper bounds for the susceptibilities which are derived from 
the reflection positivity \cite{DLS,JNFP,KLS}. For general models, some kind of boundedness about susceptibilities is needed  
for the existence of a gapless mode. In fact, the following model is very instructive: 
Let us consider two spins, ${\bf S}_1$ and ${\bf S}_2$, with $S=1/2$. The Hamiltonian is given by  
$$
H_B=-BS_1^{(3)}-S_2^{(3)},  
$$
where $B$ is the symmetry breaking field. 
When $B=0$, the spin ${\bf S}_1$ is free, and the ground state is two-fold degenerate. 
The ground state exhibits symmetry breaking for the continuous symmetry of the Hamiltonian $H_{B=0}$ in the limit $B\searrow 0$.  
However, there exists the non-zero spectral gap above the ground state. 
Of course, some of the susceptibilities are divergent. 

%%%%%%%%%%%%%%%%%%%%%%%%%%%%%%%%%%%%%%%%%%%%%%%%%%%%%%%%%%
\Section{Preliminary}
\label{Sec:Preliminary}

Before proceeding to the proof of Theorem~\ref{mainTheorem}, we prepare some tools. 

{From} (\ref{f(x)}), one has  
\begin{equation}
\label{localobservinf}
\sum_{x\in \ze^d}f(x)S_x^{(i)}=\frac{1}{(2\pi)^d}
\sum_{x\in \ze^d}\int_{-\pi}^{+\pi}dk^{(1)} \int_{-\pi}^{+\pi}dk^{(2)}\cdots \int_{-\pi}^{+\pi}dk^{(d)} \hat{f}(k)e^{ikx}S_x^{(i)}
\end{equation}
for $i=1,2,3$. In particular, 
\begin{equation}
\label{a}
a=\sum_{x\in \ze^d}f(x)S_x^{(2)}
\end{equation}
These are nothing but the spin-wave operators with the weight $\hat{f}(k)$ for the momentum $k$. 
We write  
$$
\hat{S}_k^{(i)}:=\frac{1}{\sqrt{|\Lambda|}}\sum_x e^{ikx}S_x^{(i)}
$$
for $i=1,2,3$, and  
\begin{equation}
\label{aLambda}
a_\Lambda:=\frac{1}{\sqrt{|\Lambda|}}\sum_k\hat{f}(k) \hat{S}_k^{(2)}. 
\end{equation}
{From} the symmetry of the Hamiltonian $H_{0,B}^{(\Lambda)}$, one has 
$$
\langle\Phi_{0,B}^{(\Lambda)},\hat{S}_k^{(i)}\Phi_{0,B}^{(\Lambda)}\rangle=0 
\ \ \mbox{for} \ \ i=2,3.
$$
Therefore, the reflection positivity \cite{DLS} yields the following bounds in the zero temperature limit \cite{JNFP}: 
\begin{equation}
\label{IRB}
\langle \Phi_{0,B}^{(\Lambda)},\hat{S}_{-k}^{(i)}\frac{1-P_0}{H_{0,B}^{(\Lambda)}-E_{0,B}^{(\Lambda)}}
\hat{S}_k^{(i)}\Phi_{0,B}^{(\Lambda)}\rangle\le \frac{1}{2\mathcal{E}_{k+Q}}
\end{equation}
for $i=2,3$, where $P_0$ is the projection onto the ground state $\Phi_{0,B}^{(\Lambda)}$ 
with the energy eigenvalue $E_{0,B}^{(\Lambda)}$, 
the momentum $Q$ is given by $Q=(\pi,\pi,\ldots,\pi)$, and 
$$
\mathcal{E}_k=d-\sum_{i=1}^d \cos k^{(i)}.
$$
In the following, we will often use the inequality, 
\begin{equation}
\label{doublecommu}
\langle\Phi_{0,B}^{(\Lambda)},A^\ast[H_{0,B}^{(\Lambda)}-E_{0,B}^{(\Lambda)}]A\Phi_{0,B}^{(\Lambda)}\rangle 
\le \langle \Phi_{0,B}^{(\Lambda)},[A^\ast,[H_{0,B}^{(\Lambda)},A]]\Phi_{0,B}^{(\Lambda)}\rangle,
\end{equation}
for any operator $A$. Further, the following formula is useful:  
\begin{eqnarray}
\label{tauAPhi}
\tau_{\ast g,B}^{(\Lambda)}(A)\Phi_{0,B}^{(\Lambda)}&=&
\int_{-\infty}^{+\infty}dt\; g(t) \tau_{t,B}^{(\Lambda)}(A)\Phi_{0,B}^{(\Lambda)}\nonumber\\
&=&\int_{-\infty}^{+\infty}dt\; g(t)\exp[iH_{0,B}^{(\Lambda)}t]A\exp[-iH_{0,B}^{(\Lambda)}t]\Phi_{0,B}^{(\Lambda)}\nonumber\\
&=&\int_{-\infty}^{+\infty}dt\; g(t)\exp[i(H_{0,B}^{(\Lambda)}-E_{0,B}^{(\Lambda)})t]A\Phi_{0,B}^{(\Lambda)}\nonumber\\
&=&\hat{g}(H_{0,B}^{(\Lambda)}-E_{0,B}^{(\Lambda)})A\Phi_{0,B}^{(\Lambda)} 
\end{eqnarray}
for any local observable $A$.

%%%%%%%%%%%%%%%%%%%%%%%%%%%%%%%%%%%%%%%%%%%%%%%%%%%%%%%%%%%
\Section{Excitation energies in the infinite-volume limit} 
\label{Sec:Excitations}

In this section, we recall the standard thermodynamic formalism \cite{AL,KomaS} to deal with excitation energies above 
infinite-volume ground states. If a reader is familiar with the formalism or is not interested in 
the mathematical justification, then one can skip this section. 

Since the function $\hat{f}(k)$ in  (\ref{f(x)}) is infinitely differentiable, 
the function $f(x)$ in (\ref{localobservinf}) decays by power law with distance and with a desired large power. 
Therefore, we can approximate the function $f(x)$ by a function $f^{(n)}(x)$ with a compact support 
such that $f^{(n)}(x)\rightarrow f(x)$ as $n\rightarrow \infty$ for all $x\in\ze^d$. 
Then, the Fourier transform $\hat{f}^{(n)}(k)$ of $f^{(n)}(x)$ converges to $\hat{f}(k)$ as $n\rightarrow\infty$. 
In the following, we assume ${\rm supp}\; f^{(n)}\subset\Lambda$. 
Namely, we take the limit $n\rightarrow\infty$ after taking the limit $\Lambda\nearrow\ze^d$.  

Our aim of this section is to prove the following theorem: 

\begin{theorem}
\label{subTheorem}
The following relation is valid:  
\begin{equation}
\label{Sec:aim0}
\lim_{\Gamma\nearrow\ze^d}\frac{\omega_0([\tau_{\ast g,0}(a)]^\ast[H_{0,0}^{(\Gamma)},\tau_{\ast g,0}(a)])}
{\omega_0([\tau_{\ast g,0}(a)]^\ast\tau_{\ast g,0}(a))}
=\lim_{B\searrow 0}\lim_{\Lambda\nearrow\ze^d}
\frac{\langle\Phi_{0,B}^{(\Lambda)},[\tau_{\ast g,B}^{(\Lambda)}(a_\Lambda)]^\ast
[H_{0,B}^{(\Lambda)},\tau_{\ast g,B}^{(\Lambda)}(a_\Lambda)]
\Phi_{0,B}^{(\Lambda)}\rangle}{\langle\Phi_{0,B}^{(\Lambda)},[\tau_{\ast g,B}^{(\Lambda)}(a_\Lambda)]^\ast
\tau_{\ast g,B}^{(\Lambda)}(a_\Lambda)\Phi_{0,B}^{(\Lambda)}\rangle},
\end{equation}
where the observable $a_\Lambda$ for the finite lattice $\Lambda$ is given by 
\begin{equation}
\label{aLambdadef}
a_\Lambda=\frac{1}{\sqrt{|\Lambda|}}\sum_k \hat{f}(k) \hat{S}_k^{(2)}\quad \mbox{with}\quad  
\hat{S}_k^{(2)}=\frac{1}{\sqrt{|\Lambda|}}\sum_{x\in\Lambda} e^{ikx}S_x^{(2)}.
\end{equation}
Here, if necessary, the limits are taken to be a subsequence. 
\end{theorem}

In passing, we stress the following: Write 
$$
f_\Lambda(x):=\frac{1}{|\Lambda|}\sum_k \hat{f}(k)e^{ikx}.
$$
Then, this function $f_\Lambda(x)$ is not necessarily equal to $f(x)$ which is given by (\ref{f(x)}). 
Of course, one has $f_\Lambda(x)\rightarrow f(x)$ as $\Lambda\rightarrow \ze^d$. 
\medskip

We first prove the following proposition: 

\begin{pro}
\label{Pro}
The following relation holds: 
\begin{eqnarray}
\label{Sec:aim}
& &\lim_{\Gamma\nearrow\ze^d}\frac{\omega_0([\tau_{\ast g,0}(a^{(n)})]^\ast[H_{0,0}^{(\Gamma)},\tau_{\ast g,0}(a^{(n)})])}
{\omega_0([\tau_{\ast g,0}(a^{(n)})]^\ast\tau_{\ast g,0}(a^{(n)}))}\nonumber\\
&=&\lim_{B\searrow 0}\lim_{\Lambda\nearrow\ze^d}
\frac{\langle\Phi_{0,B}^{(\Lambda)},[\tau_{\ast g,B}^{(\Lambda)}(a^{(n)})]^\ast
[H_{0,B}^{(\Lambda)},\tau_{\ast g,B}^{(\Lambda)}(a^{(n)})]
\Phi_{0,B}^{(\Lambda)}\rangle}{\langle\Phi_{0,B}^{(\Lambda)},[\tau_{\ast g,B}^{(\Lambda)}(a^{(n)})]^\ast
\tau_{\ast g,B}^{(\Lambda)}(a^{(n)})\Phi_{0,B}^{(\Lambda)}\rangle}. 
\end{eqnarray}
\end{pro}

After proving this proposition, we will take the limit $n\nearrow\infty$. 
\bigskip

\begin{proof}{Proof of Proposition~\ref{Pro}}
The observable $a^{(n)}$ is given by (\ref{a0}), i.e.,    
\begin{equation}
\label{a(n)}
a^{(n)}=\sum_x f^{(n)}(x) S_x^{(2)}  
\end{equation}
with the above approximate function $f^{(n)}$ with a compact support. In the following, we require 
\begin{equation}
\label{denominonzerobound}
\lim_{B\searrow 0}\lim_{\Lambda\nearrow\ze^d}\langle\Phi_{0,B}^{(\Lambda)},[\tau_{\ast g,B}^{(\Lambda)}(a^{(n)})]^\ast
\tau_{\ast g,B}^{(\Lambda)}(a^{(n)})\Phi_{0,B}^{(\Lambda)}\rangle>0 
\end{equation}
for a sufficiently large $n$. In the next Sec.~\ref{Sec:proof}, we will show  
$$
\lim_{B\searrow 0}\lim_{\Lambda\nearrow\ze^d}\langle\Phi_{0,B}^{(\Lambda)},[\tau_{\ast g,B}^{(\Lambda)}(a_\Lambda)]^\ast
\tau_{\ast g,B}^{(\Lambda)}(a_\Lambda)\Phi_{0,B}^{(\Lambda)}\rangle>0.  
$$
Therefore, we will show 
\begin{eqnarray*}
& &\lim_{n\nearrow\infty}
\lim_{B\searrow 0}\lim_{\Lambda\nearrow\ze^d}\langle\Phi_{0,B}^{(\Lambda)},[\tau_{\ast g,B}^{(\Lambda)}(a^{(n)})]^\ast
\tau_{\ast g,B}^{(\Lambda)}(a^{(n)})\Phi_{0,B}^{(\Lambda)}\rangle\\
&=&\lim_{B\searrow 0}\lim_{\Lambda\nearrow\ze^d}\langle\Phi_{0,B}^{(\Lambda)},[\tau_{\ast g,B}^{(\Lambda)}(a_\Lambda)]^\ast
\tau_{\ast g,B}^{(\Lambda)}(a_\Lambda)\Phi_{0,B}^{(\Lambda)}\rangle  
\end{eqnarray*}
in the proof of Theorem~\ref{subTheorem} below. 

Since the right-hand side of (\ref{a(n)}) is a finite sum, 
it is enough to consider the quantity, $\omega_0([\tau_{\ast g,0}(a_y^{(n)})]^\ast[H_{0,0}^{(\Gamma)},\tau_{\ast g,0}(a_x^{(n)})])$, 
for $x,y\in{\rm supp} f^{(n)}$, where we have written $a_x^{(n)}:=f^{(n)}(x) S_x^{(2)}$. 
The existence and boundedness of $\lim_{\Gamma\nearrow \ze^d}[H_{0,0}^{(\Gamma)},\tau_{\ast g,0}(a_x^{(n)})]$ 
are justified in Appendix~\ref{Sec:QuasiLocal}. 

We write 
$$
\omega_{0,B}^{(\Lambda)}(\cdots):=\langle\Phi_{0,B}^{(\Lambda)},(\cdots)\Phi_{0,B}^{(\Lambda)}\rangle, 
$$
and 
$$
\omega_{0,B}(\cdots):=\mbox{weak}^\ast\mbox{-}\lim_{\Lambda\nearrow\ze^d}\omega_{0,B}^{(\Lambda)}(\cdots). 
$$
Since $\omega_0(\cdots)$ is the weak$^\ast$ limit of $\omega_{0,B}(\cdots)$, there exists a sequence $\{B_i\}$ such that 
$B_i\rightarrow 0$ and $\omega_{0,B_i}(\cdots)\rightarrow \omega_0(\cdots)$ as $i\rightarrow \infty$. 
Therefore, by using the quasi-locality of the observable $\tau_{\ast g,0}(a_x^{(n)})$ which is proved in Appendix~\ref{Sec:QuasiLocal}, 
we have 
$$
\omega_{0,B_i}([\tau_{\ast g,0}(a_y^{(n)})]^\ast[H_{0,0}^{(\Gamma)},\tau_{\ast g,0}(a_x^{(n)})])\rightarrow 
\omega_0([\tau_{\ast g,0}(a_y^{(n)})]^\ast[H_{0,0}^{(\Gamma)},\tau_{\ast g,0}(a_x^{(n)})])\quad \mbox{as} \ \ B_i\rightarrow 0. 
$$
Further, by the continuity (\ref{contitauB}) with respect to $B$ in Appendix~\ref{Sec:continuity}, we obtain 
$$
\omega_{0,B_i}([\tau_{\ast g,B_i}(a_y^{(n)})]^\ast[H_{0,B_i}^{(\Gamma)},\tau_{\ast g,B_i}(a_x^{(n)})])\rightarrow 
\omega_0([\tau_{\ast g,0}(a_y^{(n)})]^\ast[H_{0,0}^{(\Gamma)},\tau_{\ast g,0}(a_x^{(n)})])\quad \mbox{as} \ \ B_i\rightarrow 0. 
$$
In order to handle the quantity in the left-hand side, we use \cite{BHV,BMNS,NSY} 
a local approximation $\Pi_X(\cdots)$ which is defined 
by (\ref{PiX}) in Appendix~\ref{Sec:QuasiLocal}. 
Namely, for any quasi-local operator $b$, the support of $\Pi_X(b)$ is contained in $X$. 
As a result, the quantity in the left-hand side can be written 
\begin{eqnarray*}
& &\omega_{0,B_i}([\tau_{\ast g,B_i}(a_y^{(n)})]^\ast[H_{0,B_i}^{(\Gamma)},\tau_{\ast g,B_i}(a_x^{(n)})])\\
&=&\lim_{\Lambda\nearrow \ze^d}\lim_{m\nearrow \infty}
\omega_{0,B_i}([\Pi_{X_m(y)}(\tau_{\ast g,B_i}^{(\Lambda)}(a_y^{(n)}))]^\ast[H_{0,B_i}^{(\Gamma)},
\Pi_{X_m(x)}(\tau_{\ast g,B_i}^{(\Lambda)}(a_x^{(n)}))]),
\end{eqnarray*}
where $X_m(x)$ is defined by (\ref{Xm(x)}) in Appendix~\ref{Sec:QuasiLocal}, and 
we have used (\ref{Pitaudiff}) in Appendix~\ref{Sec:QuasiLocal}. From the definition of $\omega_{0,B_i}(\cdots)$, 
there exists a sequence $\{\Lambda_j\}$ such that 
$$
\lim_{j\nearrow \infty}\omega_{0,B_i}^{(\Lambda_j)}(\cdots)=\omega_{0,B_i}(\cdots).
$$ 
Therefore, one has 
\begin{eqnarray}
\label{omega0BiPimtau}
& &\omega_{0,B_i}([\Pi_{X_m(y)}(\tau_{\ast g,B_i}^{(\Lambda)}(a_y^{(n)}))]^\ast[H_{0,B_i}^{(\Gamma)},
\Pi_{X_m(x)}(\tau_{\ast g,B_i}^{(\Lambda)}(a_x^{(n)}))])\nonumber\\
&=&\lim_{j\nearrow\infty}\omega_{0,B_i}^{(\Lambda_j)}([\Pi_{X_m(y)}(\tau_{\ast g,B_i}^{(\Lambda)}(a_y^{(n)}))]^\ast[H_{0,B_i}^{(\Gamma)},
\Pi_{X_m(x)}(\tau_{\ast g,B_i}^{(\Lambda)}(a_x^{(n)}))])\nonumber\\
&=&\lim_{j\nearrow\infty}\omega_{0,B_i}^{(\Lambda_j)}([\Pi_{X_m(y)}(\tau_{\ast g,B_i}^{(\Lambda)}(a_y^{(n)}))]^\ast
[H_{0,B_i}^{(\Lambda_j)},
\Pi_{X_m(x)}(\tau_{\ast g,B_i}^{(\Lambda)}(a_x^{(n)}))]).
\end{eqnarray}
Note that 
$$
\tau_{*g,B_i}^{(\Lambda)}(a_x^{(n)})-\Pi_{X_m(x)}(\tau_{*g,B_i}^{(\Lambda)}(a_x^{(n)}))
=\sum_{\ell=m+1}^\infty \Delta_\ell^{(\Lambda)}(a_x^{(n)})
$$
which is derived from (\ref{tauDeltasum}) with (\ref{Delta0}) and (\ref{Deltam}) in Appendix~\ref{Sec:QuasiLocal}. 
This yields  
$$
\left[H_{0,B_i}^{(\Lambda_j)},\{\tau_{*g,B_i}^{(\Lambda)}(a_x^{(n)})
-\Pi_{X_m(x)}(\tau_{\ast g,B_i}^{(\Lambda)}(a_x^{(n)}))\}\right]
=\sum_{\ell=m+1}^\infty [H_{0,B_i}^{(\Lambda_j)},\Delta_\ell^{(\Lambda)}(a_x^{(n)})]. 
$$
Using this and (\ref{Deltabound}) in Appendix~\ref{Sec:QuasiLocal}, 
we can replace $\Pi_{X_m(x)}(\tau_{\ast g,B_i}^{(\Lambda)}(a_x^{(n)}))$ 
by $\tau_{\ast g,B_i}^{(\Lambda_j)}(a_x^{(n)})$ with a small error 
in the right-hand side of (\ref{omega0BiPimtau}) with sufficiently large $m$ and $\Lambda$. 
Then, the above right-hand side has the desired form, 
$$
\omega_{0,B_i}^{(\Lambda_j)}([\tau_{\ast g,B_i}^{(\Lambda_j)}(a_y^{(n)})]^\ast
[H_{0,B_i}^{(\Lambda_j)},\tau_{\ast g,B_i}^{(\Lambda_j)}(a_x^{(n)})]).
$$
This is a component of the numerator in the right-hand side of (\ref{Sec:aim}). Thus, we have 
\begin{eqnarray}
\label{numerator}
& &\lim_{\Gamma\nearrow\ze^d}\omega_0([\tau_{*g,0}(a^{(n)})]^\ast[H_{0,0}^{(\Gamma)},\tau_{*g,0}(a^{(n)})])\nonumber\\
&=&\lim_{B_i\searrow 0}\lim_{\Lambda_j\nearrow\ze^d}\langle\Phi_{0,B_i}^{(\Lambda_j)}, 
[\tau_{*g,B_i}^{(\Lambda_j)}(a^{(n)})]^\ast[H_{0,B_i}^{(\Lambda_j)},\tau_{*g,B_i}^{(\Lambda_j)}(a^{(n)})]
\Phi_{0,B_i}^{(\Lambda_j)}\rangle
\end{eqnarray}
Clearly, the dominator can be treated in the same way. Namely, 
\begin{equation}
\label{denominator}
\lim_{\Gamma\nearrow\ze^d}\omega_0([\tau_{*g,0}(a^{(n)})]^\ast \tau_{*g,0}(a^{(n)}))
=\lim_{B_i\searrow 0}\lim_{\Lambda_j\nearrow\ze^d}\langle\Phi_{0,B_i}^{(\Lambda_j)}, 
[\tau_{*g,B_i}^{(\Lambda_j)}(a^{(n)})]^\ast \tau_{*g,B_i}^{(\Lambda_j)}(a^{(n)})\Phi_{0,B_i}^{(\Lambda_j)}\rangle.
\end{equation}
Thus, the equality (\ref{Sec:aim}) has been proved, provided 
that the bound (\ref{denominonzerobound}) holds for a sufficiently large $n$. The bound (\ref{denominonzerobound}) is 
justified in the proof of Theorem~\ref{subTheorem} below. 
\end{proof} 

\begin{proof}{Proof of Theorem~\ref{subTheorem}}
Next, we take the limit $n\nearrow\infty$. We choose the function $f^{(n)}(x)$ so that it satisfies 
\begin{equation}
\label{fnxcondition}
f(x)-f^{(n)}(x)=0 \quad \mbox{for \ } |x|\le \ell_n \quad \mbox{and} \quad 
f^{(n)}(x)=0 \quad \mbox{for \ } |x|>\ell_n, 
\end{equation}
where the scale length $\ell_n$ of the support $f^{(n)}(x)$ satisfies $\ell_n\rightarrow\infty$ as $n\rightarrow \infty$. 
By the assumption on the function $\hat{f}(k)$, we can find a positive constant $\mathcal{C}_1$ and 
a sufficiently large exponent $\eta$ such that  
\begin{equation}
\label{fdiffdecay}
|f(x)|=|f(x)-f^{(n)}(x)|\le \mathcal{C}_1|x|^{-\eta} \quad \mbox{for \ } |x|>\ell_n.  
\end{equation}
Note that 
$$
\hat{f}(k)=\sum_x f(x)e^{-ikx}
$$
and 
\begin{equation}
\label{hatf(n)k}
\hat{f}^{(n)}(k):=\sum_x f^{(n)}(x)e^{-ikx}.
\end{equation}
Combining these with (\ref{fnxcondition}) and (\ref{fdiffdecay}), one has 
\begin{equation}
\label{hatfkdiff}
|\hat{f}(k)-\hat{f}^{(n)}(k)|\le \sum_{x: |x|>\ell_n}|f(x)-f^{(n)}(x)|\le \mathcal{C}_1 \sum_{x: |x|>\ell_n}|x|^{-\eta}
\le \mathcal{C}_2 \ell_n^{-\eta+d}, 
\end{equation}
where $\mathcal{C}_2$ is a positive constant, and we have chosen $\eta$ to satisfy $\eta-d>0$.  

Let us first consider the left-hand side of (\ref{numerator}). 
{From} (\ref{a}) and (\ref{a(n)}), one has 
\begin{equation}
a^{(n)}=a+\sum_x[f^{(n)}(x)-f(x)]S_x^{(2)}. 
\end{equation}
This yields 
$$
[H_{0,0}^{(\Gamma)}, \tau_{*g,0}(a^{(n)})]
=[H_{0,0}^{(\Gamma)}, \tau_{*g,0}(a)]+\sum_x [H_{0,0}^{(\Gamma)},\tau_{*g,0}(S_x^{(2)})]\cdot[f^{(n)}(x)-f(x)]. 
$$
By applying the local approximation decomposition to the observable $\tau_{*g,0}(S_x^{(2)})$, we have 
$$
\Vert[H_{0,0}^{(\Gamma)},\tau_{*g,0}(S_x^{(2)})]\Vert\le\mathcal{C}_3
$$
with a positive constant $\mathcal{C}_3$ in the same way as in Appendix~\ref{Sec:QuasiLocal}. 
(More precisely, we use the decomposition (\ref{tauDeltasum}) in the infinite-volume limit $\Lambda\nearrow\ze^d$.) 
Combining these with (\ref{fnxcondition}) and (\ref{fdiffdecay}), we obtain 
\begin{eqnarray}
\left\Vert[H_{0,0}^{(\Gamma)}, \tau_{*g,0}(a^{(n)})]-[H_{0,0}^{(\Gamma)}, \tau_{*g,0}(a)]\right\Vert
&\le&\sum_x \left\Vert[H_{0,0}^{(\Gamma)},\tau_{*g,0}(S_x^{(2)})]\right\Vert\cdot|f^{(n)}(x)-f(x)|\nonumber\\
&\le&\mathcal{C}_1\mathcal{C}_3\sum_{x:|x|>\ell_n} |x|^{-\eta} \le \mathcal{C}_4\ell_n^{-\eta+d}
\end{eqnarray}
with a positive constant $\mathcal{C}_4$. Therefore, we have 
\begin{equation}
\lim_{n\nearrow\infty}\lim_{\Gamma\nearrow\ze^d}\omega_0([\tau_{*g,0}(a^{(n)})]^\ast[H_{0,0}^{(\Gamma)}, \tau_{*g,0}(a^{(n)})])
=\lim_{\Gamma\nearrow\ze^d}\omega_0([\tau_{*g,0}(a)]^\ast[H_{0,0}^{(\Gamma)}, \tau_{*g,0}(a)])
\end{equation} 
and 
\begin{equation}
\lim_{n\nearrow\infty}\lim_{\Gamma\nearrow\ze^d}\omega_0([\tau_{*g,0}(a^{(n)})]^\ast \tau_{*g,0}(a^{(n)}))
=\lim_{\Gamma\nearrow\ze^d}\omega_0([\tau_{*g,0}(a)]^\ast \tau_{*g,0}(a)). 
\end{equation} 
 
Next consider the right-hand side of (\ref{numerator}). 
{From} (\ref{a(n)}) and (\ref{hatf(n)k}), the observable $a^{(n)}$ is written 
$$
a^{(n)}=\frac{1}{\sqrt{|\Lambda|}}\sum_k \hat{f}^{(n)}(k)\hat{S}_k^{(2)}.
$$
Using this expression and (\ref{tauAPhi}), we have 
\begin{eqnarray}
& &\langle\Phi_{0,B}^{(\Lambda)},[\tau_{\ast g,B}^{(\Lambda)}(a^{(n)})]^\ast[H_{0,B}^{(\Lambda)},
\tau_{\ast g,B}^{(\Lambda)}(a^{(n)})]
\Phi_{0,B}^{(\Lambda)}\rangle \nonumber\\
&=& \langle\Phi_{0,B}^{(\Lambda)},[a^{(n)}]^\ast [H_{0,B}^{(\Lambda)}-E_{0,B}^{(\Lambda)}]
[\hat{g}(H_{0,B}^{(\Lambda)}-E_{0,B}^{(\Lambda)})]^2a^{(n)}\Phi_{0,B}^{(\Lambda)}\rangle\nonumber\\
&=&\frac{1}{|\Lambda|}\sum_{k,k'}\hat{f}^{(n)}(k)^\ast \hat{f}^{(n)}(k')
\langle\Phi_{0,B}^{(\Lambda)},\hat{S}_{-k}^{(2)} [H_{0,B}^{(\Lambda)}-E_{0,B}^{(\Lambda)}]
[\hat{g}(H_{0,B}^{(\Lambda)}-E_{0,B}^{(\Lambda)})]^2\hat{S}_{k'}^{(2)}\Phi_{0,B}^{(\Lambda)}\rangle\nonumber\\
&=&\frac{1}{|\Lambda|}\sum_{k}|\hat{f}^{(n)}(k)|^2
\langle\Phi_{0,B}^{(\Lambda)},\hat{S}_{-k}^{(2)} [H_{0,B}^{(\Lambda)}-E_{0,B}^{(\Lambda)}]
[\hat{g}(H_{0,B}^{(\Lambda)}-E_{0,B}^{(\Lambda)})]^2\hat{S}_{k}^{(2)}\Phi_{0,B}^{(\Lambda)}\rangle,
\end{eqnarray} 
where $E_{0,B}^{(\Lambda)}$ is the ground state energy of the Hamiltonian $H_{0,B}^{(\Lambda)}$, 
and we have used the momentum conservation of the Hamiltonian $\tilde{H}_{0,B}^{(\Lambda)}$ of (\ref{tildeham}). 
Similarly, from (\ref{aLambdadef}), one has 
\begin{eqnarray}
& &\langle\Phi_{0,B}^{(\Lambda)},[\tau_{\ast g,B}^{(\Lambda)}(a_\Lambda)]^\ast[H_{0,B}^{(\Lambda)},
\tau_{\ast g,B}^{(\Lambda)}(a_\Lambda)]
\Phi_{0,B}^{(\Lambda)}\rangle \nonumber\\
&=&\frac{1}{|\Lambda|}\sum_{k}|\hat{f}(k)|^2
\langle\Phi_{0,B}^{(\Lambda)},\hat{S}_{-k}^{(2)} [H_{0,B}^{(\Lambda)}-E_{0,B}^{(\Lambda)}]
[\hat{g}(H_{0,B}^{(\Lambda)}-E_{0,B}^{(\Lambda)})]^2\hat{S}_{k}^{(2)}\Phi_{0,B}^{(\Lambda)}\rangle.
\end{eqnarray} 
Combining these with (\ref{hatfkdiff}), we obtain 
\begin{eqnarray}
& &\left|\langle\Phi_{0,B}^{(\Lambda)},[\tau_{\ast g,B}^{(\Lambda)}(a^{(n)})]^\ast[H_{0,B}^{(\Lambda)},
\tau_{\ast g,B}^{(\Lambda)}(a^{(n)})]\Phi_{0,B}^{(\Lambda)}\rangle 
- \langle\Phi_{0,B}^{(\Lambda)},[\tau_{\ast g,B}^{(\Lambda)}(a_\Lambda)]^\ast[H_{0,B}^{(\Lambda)},
\tau_{\ast g,B}^{(\Lambda)}(a_\Lambda)]
\Phi_{0,B}^{(\Lambda)}\rangle\right|\nonumber\\
& &\le \mathcal{C}_5 \ell_n^{-\eta+d} \frac{1}{|\Lambda|}\sum_{k}
\langle\Phi_{0,B}^{(\Lambda)},\hat{S}_{-k}^{(2)} [H_{0,B}^{(\Lambda)}-E_{0,B}^{(\Lambda)}]
[\hat{g}(H_{0,B}^{(\Lambda)}-E_{0,B}^{(\Lambda)})]^2\hat{S}_{k}^{(2)}\Phi_{0,B}^{(\Lambda)}\rangle\nonumber\\
& &\le\mathcal{C}_5 \ell_n^{-\eta+d} \frac{1}{|\Lambda|}\sum_{k}
\langle\Phi_{0,B}^{(\Lambda)},\hat{S}_{-k}^{(2)} [H_{0,B}^{(\Lambda)}-E_{0,B}^{(\Lambda)}]
\hat{S}_{k}^{(2)}\Phi_{0,B}^{(\Lambda)}\rangle\nonumber\\
& &\le \mathcal{C}_5 \ell_n^{-\eta+d} \frac{1}{|\Lambda|}\sum_{k}
\langle\Phi_{0,B}^{(\Lambda)},[\hat{S}_{-k}^{(2)}, [H_{0,B}^{(\Lambda)},\hat{S}_{k}^{(2)}]]\Phi_{0,B}^{(\Lambda)}\rangle\nonumber\\
& &\le \mathcal{C}_5 \ell_n^{-\eta+d} \frac{1}{|\Lambda|}\sum_{k}(4S^2\mathcal{E}_k + B S), 
\end{eqnarray}
where 
$$
\mathcal{E}_k=d-\sum_{i=1}^d \cos k^{(i)}.
$$
Clearly, the right-hand side is vanishing in the limit $n\nearrow\infty$ because $\eta>d$. Similarly, we have 
\begin{eqnarray}
\label{diffdomi}
& &\left|\langle\Phi_{0,B}^{(\Lambda)},[\tau_{\ast g,B}^{(\Lambda)}(a^{(n)})]^\ast
\tau_{\ast g,B}^{(\Lambda)}(a^{(n)})\Phi_{0,B}^{(\Lambda)}\rangle 
- \langle\Phi_{0,B}^{(\Lambda)},[\tau_{\ast g,B}^{(\Lambda)}(a_\Lambda)]^\ast
\tau_{\ast g,B}^{(\Lambda)}(a_\Lambda)\Phi_{0,B}^{(\Lambda)}\rangle\right|\nonumber\\
&\le&\mathcal{C}_5 \ell_n^{-\eta+d} \frac{1}{|\Lambda|}\sum_{k}
\langle\Phi_{0,B}^{(\Lambda)},\hat{S}_{-k}^{(2)} (1-P_0)\hat{S}_{k}^{(2)}\Phi_{0,B}^{(\Lambda)}\rangle,
\end{eqnarray}
where $P_0$ is the projection onto the ground state $\Phi_{0,B}^{(\Lambda)}$. 
By using the bound (\ref{IRB}) and the Schwarz inequality, the summand in the right-hand side can be estimated as follows: 
\begin{eqnarray}
\langle\Phi_{0,B}^{(\Lambda)},\hat{S}_{-k}^{(2)} (1-P_0)\hat{S}_{k}^{(2)}\Phi_{0,B}^{(\Lambda)}\rangle&=&
\Big\langle\Phi_{0,B}^{(\Lambda)},\hat{S}_{-k}^{(2)} \frac{1-P_0}{\sqrt{H_{0,B}^{(\Lambda)}-E_{0,B}^{(\Lambda)}}}
\sqrt{H_{0,B}^{(\Lambda)}-E_{0,B}^{(\Lambda)}}\hat{S}_{k}^{(2)}\Phi_{0,B}^{(\Lambda)}\Big\rangle\nonumber\\
&\le&\sqrt{\Big\langle\Phi_{0,B}^{(\Lambda)},\hat{S}_{-k}^{(2)} \frac{1-P_0}{{H_{0,B}^{(\Lambda)}-E_{0,B}^{(\Lambda)}}}
\hat{S}_{k}^{(2)}\Phi_{0,B}^{(\Lambda)}\Big\rangle}\nonumber\\
&\times&\sqrt{\Big\langle\Phi_{0,B}^{(\Lambda)},\hat{S}_{-k}^{(2)}
({H_{0,B}^{(\Lambda)}-E_{0,B}^{(\Lambda)}})\hat{S}_{k}^{(2)}\Phi_{0,B}^{(\Lambda)}\Big\rangle}\nonumber\\
&\le& \sqrt{\frac{4S^2\mathcal{E}_k + B S}{2\mathcal{E}_{k+Q}}}. 
\end{eqnarray}
This bound yields 
\begin{eqnarray} 
\frac{1}{|\Lambda|}\sum_{k}
\langle\Phi_{0,B}^{(\Lambda)},\hat{S}_{-k}^{(2)} (1-P_0)\hat{S}_{k}^{(2)}\Phi_{0,B}^{(\Lambda)}\rangle
&=&\frac{1}{|\Lambda|}\langle\Phi_{0,B}^{(\Lambda)},\hat{S}_{-Q}^{(2)} (1-P_0)\hat{S}_{Q}^{(2)}\Phi_{0,B}^{(\Lambda)}\rangle\nonumber\\
&+&\frac{1}{|\Lambda|}\sum_{k\ne Q}
\langle\Phi_{0,B}^{(\Lambda)},\hat{S}_{-k}^{(2)} (1-P_0)\hat{S}_{k}^{(2)}\Phi_{0,B}^{(\Lambda)}\rangle\nonumber\\
&\le& \frac{1}{|\Lambda|}\langle\Phi_{0,B}^{(\Lambda)},\hat{S}_{-Q}^{(2)}\hat{S}_{Q}^{(2)}\Phi_{0,B}^{(\Lambda)}\rangle
+\frac{1}{|\Lambda|}\sum_{k\ne Q}\sqrt{\frac{4S^2\mathcal{E}_k + B S}{2\mathcal{E}_{k+Q}}}.\nonumber\\
\end{eqnarray}
This right-hand side is bounded by some constant because $d\ge 2$. Therefore, the right-hand side of (\ref{diffdomi}) 
is vanishing in the limit $n\nearrow\infty$. As mentioned above, this result implies that 
the bound (\ref{denominonzerobound}) is justified, too. 
\end{proof}

%%%%%%%%%%%%%%%%%%%%%%%%%%%%%%%%%%%%%%%%%%%%%%%%%%%%%%%%%%%%%%%%%%%%
\Section{Proof of Theorem~\ref{mainTheorem}}
\label{Sec:proof}

By relying on Theorem~\ref{subTheorem}, we give a proof of Theorem~\ref{mainTheorem}. 
More precisely, we will obtain the upper and lower bounds for the right-hand side of (\ref{Sec:aim0}). 

{From} (\ref{tauAPhi}) and the expression (\ref{aLambda}) of the observable $a_\Lambda$, we have 
\begin{eqnarray}
\label{numerbound}
& &\langle\Phi_{0,B}^{(\Lambda)},[\tau_{\ast g,B}^{(\Lambda)}(a_\Lambda)]^\ast[H_{0,B}^{(\Lambda)},
\tau_{\ast g,B}^{(\Lambda)}(a_\Lambda)]
\Phi_{0,B}^{(\Lambda)}\rangle \nonumber\\
&=& \langle\Phi_{0,B}^{(\Lambda)},a_\Lambda^\ast [H_{0,B}^{(\Lambda)}-E_{0,B}^{(\Lambda)}]
[\hat{g}(H_{0,B}^{(\Lambda)}-E_{0,B}^{(\Lambda)})]^2a_\Lambda\Phi_{0,B}^{(\Lambda)}\rangle\nonumber\\
&=&\frac{1}{|\Lambda|}\sum_{k,k'}\hat{f}(k)^\ast \hat{f}(k')
\langle\Phi_{0,B}^{(\Lambda)},\hat{S}_{-k}^{(2)} [H_{0,B}^{(\Lambda)}-E_{0,B}^{(\Lambda)}]
[\hat{g}(H_{0,B}^{(\Lambda)}-E_{0,B}^{(\Lambda)})]^2\hat{S}_{k'}^{(2)}\Phi_{0,B}^{(\Lambda)}\rangle\nonumber\\
&=&\frac{1}{|\Lambda|}\sum_{k}|\hat{f}(k)|^2
\langle\Phi_{0,B}^{(\Lambda)},\hat{S}_{-k}^{(2)} [H_{0,B}^{(\Lambda)}-E_{0,B}^{(\Lambda)}]
[\hat{g}(H_{0,B}^{(\Lambda)}-E_{0,B}^{(\Lambda)})]^2\hat{S}_{k}^{(2)}\Phi_{0,B}^{(\Lambda)}\rangle\nonumber\\
&\le& \frac{1}{|\Lambda|}\sum_{k}|\hat{f}(k)|^2
\langle\Phi_{0,B}^{(\Lambda)},\hat{S}_{-k}^{(2)} [H_{0,B}^{(\Lambda)}-E_{0,B}^{(\Lambda)}]
\hat{S}_{k}^{(2)}\Phi_{0,B}^{(\Lambda)}\rangle\nonumber\\
&\le&\frac{1}{|\Lambda|}\sum_{k}|\hat{f}(k)|^2
\langle\Phi_{0,B}^{(\Lambda)},[\hat{S}_{-k}^{(2)}, [H_{0,B}^{(\Lambda)},\hat{S}_{k}^{(2)}]]\Phi_{0,B}^{(\Lambda)}\rangle\nonumber\\
&\le&\frac{1}{|\Lambda|}\sum_{k}|\hat{f}(k)|^2[ 4S^2\mathcal{E}_k + B S], 
\end{eqnarray}
where we have used the momentum conservation of the Hamiltonian $\tilde{H}_{0,B}^{(\Lambda)}$ of 
(\ref{tildeham}) and the inequality (\ref{doublecommu}).  

Similarly, one has 
\begin{equation}
\label{denomi}
\langle\Phi_{0,B}^{(\Lambda)},[\tau_{\ast g,B}^{(\Lambda)}(a_\Lambda)]^\ast
\tau_{\ast g,B}^{(\Lambda)}(a_\Lambda)\Phi_{0,B}^{(\Lambda)}\rangle
=\frac{1}{|\Lambda|}\sum_{k}|\hat{f}(k)|^2
\langle\Phi_{0,B}^{(\Lambda)},\hat{S}_{-k}^{(2)} 
[\hat{g}(H_{0,B}^{(\Lambda)}-E_{0,B}^{(\Lambda)})]^2\hat{S}_{k}^{(2)}\Phi_{0,B}^{(\Lambda)}\rangle. 
\end{equation}
We recall the assumption (\ref{supphatf}) on the function $\hat{f}(k)$, i.e., 
${\rm supp}\; \hat{f}\subset \{k\;|\; {|p|}/{2}\le |k|\le |p|\}$ with a small $|p|>0$. 
Therefore, in the following, we will consider only the momenta $k$  
that satisfy the condition ${|p|}/{2}\le |k|\le |p|$ with a small $|p|>0$.  
Clearly, the other contributions are vanishing in the sum of the right-hand side of (\ref{denomi}). 
The crucial point is that the two momenta, $k=0$ and $k=Q$, may cause a divergence for the upper bounds of the susceptibilities. 

In order to estimate the right-hand side of (\ref{denomi}), 
we consider the staggered magnetization $m_B^{(\Lambda)}$ of (\ref{mBLambda}). 
One has the expression, 
\begin{equation}
m_B^{(\Lambda)}=-i \langle\Phi_{0,B}^{(\Lambda)},[\hat{S}_{-k}^{(2)},\hat{S}_{Q+k}^{(3)}]\Phi_{0,B}^{(\Lambda)}\rangle.
\end{equation}
The quantity in the right-hand side can be written 
\begin{equation}
\label{msexpcommu}
\langle\Phi_{0,B}^{(\Lambda)},[\hat{S}_{-k}^{(2)},\hat{S}_{Q+k}^{(3)}]\Phi_{0,B}^{(\Lambda)}\rangle
=\langle\Phi_{0,B}^{(\Lambda)},\hat{S}_{-k}^{(2)}(1-P_0)\hat{S}_{Q+k}^{(3)}\Phi_{0,B}^{(\Lambda)}\rangle
-\langle\Phi_{0,B}^{(\Lambda)},\hat{S}_{Q+k}^{(3)}(1-P_0)\hat{S}_{-k}^{(2)}\Phi_{0,B}^{(\Lambda)}\rangle.
\end{equation}
In order to evaluate the two terms in the right-hand side, we decompose the projection $1-P_0$ into two projections, 
$P(0,2\epsilon]$ onto the subspace of the energy interval $(0,2\epsilon]$ 
and $P(2\epsilon,\infty)$ for the interval $(2\epsilon,\infty)$, i.e.,  
$$
1-P_0=P(0,2\epsilon]+P(2\epsilon,\infty).
$$
Then, we have 
\begin{eqnarray}
\label{contrib02epsilon}
|\langle\Phi_{0,B}^{(\Lambda)},\hat{S}_{-k}^{(2)}P(0,2\epsilon]\hat{S}_{Q+k}^{(3)}\Phi_{0,B}^{(\Lambda)}\rangle|
&\le&\sqrt{\langle\Phi_{0,B}^{(\Lambda)},\hat{S}_{-k}^{(2)}P(0,2\epsilon]\hat{S}_{k}^{(2)}\Phi_{0,B}^{(\Lambda)}\rangle}\nonumber\\ 
&\times&\sqrt{\langle\Phi_{0,B}^{(\Lambda)},\hat{S}_{Q-k}^{(3)}P(0,2\epsilon]\hat{S}_{Q+k}^{(3)}\Phi_{0,B}^{(\Lambda)}\rangle}\nonumber\\
&\le&\sqrt{\langle\Phi_{0,B}^{(\Lambda)},\hat{S}_{-k}^{(2)}\frac{P(0,2\epsilon]}{H_{0,B}^{(\Lambda)}-E_{0,B}^{(\Lambda)}}
[H_{0,B}^{(\Lambda)}-E_{0,B}^{(\Lambda)}]
\hat{S}_{k}^{(2)}\Phi_{0,B}^{(\Lambda)}\rangle}\nonumber\\
&\times&\sqrt{\langle\Phi_{0,B}^{(\Lambda)},\hat{S}_{Q-k}^{(3)}\frac{P(0,2\epsilon]}{H_{0,B}^{(\Lambda)}-E_{0,B}^{(\Lambda)}}
[H_{0,B}^{(\Lambda)}-E_{0,B}^{(\Lambda)}]
\hat{S}_{Q+k}^{(3)}\Phi_{0,B}^{(\Lambda)}\rangle}\nonumber\\
&\le& 2\epsilon \sqrt{\langle\Phi_{0,B}^{(\Lambda)},\hat{S}_{-k}^{(2)}\frac{1-P_0}{H_{0,B}^{(\Lambda)}-E_{0,B}^{(\Lambda)}}
\hat{S}_{k}^{(2)}\Phi_{0,B}^{(\Lambda)}\rangle}\nonumber\\
&\times&\sqrt{\langle\Phi_{0,B}^{(\Lambda)},\hat{S}_{Q-k}^{(3)}\frac{1-P_0}{H_{0,B}^{(\Lambda)}-E_{0,B}^{(\Lambda)}}
\hat{S}_{Q+k}^{(3)}\Phi_{0,B}^{(\Lambda)}\rangle}\nonumber\\
&\le& \frac{\epsilon}{\sqrt{\mathcal{E}_{k+Q}\mathcal{E}_k}},
\end{eqnarray} 
where we have used the bounds (\ref{IRB}). 

Similarly, one has 
\begin{eqnarray}
|\langle\Phi_{0,B}^{(\Lambda)},\hat{S}_{-k}^{(2)}P(2\epsilon,\infty)\hat{S}_{Q+k}^{(3)}\Phi_{0,B}^{(\Lambda)}\rangle|
&\le&\sqrt{\langle\Phi_{0,B}^{(\Lambda)},\hat{S}_{-k}^{(2)}P(2\epsilon,\infty)\hat{S}_{k}^{(2)}\Phi_{0,B}^{(\Lambda)}\rangle}\nonumber\\ 
&\times&\sqrt{\langle\Phi_{0,B}^{(\Lambda)},\hat{S}_{Q-k}^{(3)}P(2\epsilon,\infty)\hat{S}_{Q+k}^{(3)}\Phi_{0,B}^{(\Lambda)}\rangle}.
\end{eqnarray}
Note that 
\begin{eqnarray}
\langle\Phi_{0,B}^{(\Lambda)},\hat{S}_{-k}^{(2)}P(2\epsilon,\infty)\hat{S}_{k}^{(2)}\Phi_{0,B}^{(\Lambda)}\rangle
&\le& \langle\Phi_{0,B}^{(\Lambda)},\hat{S}_{-k}^{(2)}[\hat{g}(H_{0,B}^{(\Lambda)}-E_{0,B}^{(\Lambda)})]^2\hat{S}_{k}^{(2)}
\Phi_{0,B}^{(\Lambda)}\rangle\nonumber\\
&+&\langle\Phi_{0,B}^{(\Lambda)},\hat{S}_{-k}^{(2)}P[\gamma-\Delta\gamma,\infty)\hat{S}_{k}^{(2)}\Phi_{0,B}^{(\Lambda)}\rangle\nonumber\\
&\le& \langle\Phi_{0,B}^{(\Lambda)},\hat{S}_{-k}^{(2)}[\hat{g}(H_{0,B}^{(\Lambda)}-E_{0,B}^{(\Lambda)})]^2\hat{S}_{k}^{(2)}
\Phi_{0,B}^{(\Lambda)}\rangle\nonumber\\
&+&\frac{1}{\gamma-\Delta\gamma}\langle\Phi_{0,B}^{(\Lambda)},\hat{S}_{-k}^{(2)}[H_{0,B}^{(\Lambda)}-E_{0,B}^{(\Lambda)}]
\hat{S}_{k}^{(2)}\Phi_{0,B}^{(\Lambda)}\rangle\nonumber\\
&\le& \langle\Phi_{0,B}^{(\Lambda)},\hat{S}_{-k}^{(2)}[\hat{g}(H_{0,B}^{(\Lambda)}-E_{0,B}^{(\Lambda)})]^2\hat{S}_{k}^{(2)}
\Phi_{0,B}^{(\Lambda)}\rangle\nonumber\\
&+&\frac{1}{\gamma-\Delta\gamma}\langle\Phi_{0,B}^{(\Lambda)},[\hat{S}_{-k}^{(2)},[H_{0,B}^{(\Lambda)},\hat{S}_{k}^{(2)}]]
\Phi_{0,B}^{(\Lambda)}\rangle\nonumber\\
&\le& \langle\Phi_{0,B}^{(\Lambda)},\hat{S}_{-k}^{(2)}[\hat{g}(H_{0,B}^{(\Lambda)}-E_{0,B}^{(\Lambda)})]^2\hat{S}_{k}^{(2)}
\Phi_{0,B}^{(\Lambda)}\rangle\nonumber\\
&+&\frac{1}{\gamma-\Delta\gamma}[ 4S^2\mathcal{E}_k + B S],
\end{eqnarray}
where $P[\gamma-\Delta\gamma,\infty)$ is the projection for the energy interval $[\gamma-\Delta\gamma,\infty)$, 
and we have used the assumption on the function $\hat{g}$. 
Further, 
\begin{eqnarray}
\langle\Phi_{0,B}^{(\Lambda)},\hat{S}_{Q-k}^{(3)}P(2\epsilon,\infty)\hat{S}_{Q+k}^{(3)}\Phi_{0,B}^{(\Lambda)}\rangle
&\le&\sqrt{\langle\Phi_{0,B}^{(\Lambda)},\hat{S}_{Q-k}^{(3)}[H_{0,B}^{(\Lambda)}-E_{0,B}^{(\Lambda)}]\hat{S}_{Q+k}^{(3)}
\Phi_{0,B}^{(\Lambda)}\rangle}\nonumber\\
&\times&\sqrt{\langle\Phi_{0,B}^{(\Lambda)},\hat{S}_{Q-k}^{(3)}\frac{1}{H_{0,B}^{(\Lambda)}-E_{0,B}^{(\Lambda)}}\hat{S}_{Q+k}^{(3)}
\Phi_{0,B}^{(\Lambda)}\rangle}\nonumber\\
&\le&\sqrt{\langle\Phi_{0,B}^{(\Lambda)},[\hat{S}_{Q-k}^{(3)},[H_{0,B}^{(\Lambda)},\hat{S}_{Q+k}^{(3)}]
\Phi_{0,B}^{(\Lambda)}\rangle}\times \frac{1}{\sqrt{2\mathcal{E}_k}}\nonumber\\
&\le& \sqrt{\frac{4S^2\mathcal{E}_{k+Q} + B S}{2\mathcal{E}_k}}. 
\end{eqnarray}
By putting all together, one has 
\begin{eqnarray}
\label{msexpcommu1b}
& &\left|\langle\Phi_{0,B}^{(\Lambda)},\hat{S}_{-k}^{(2)}(1-P_0)\hat{S}_{Q+k}^{(3)}\Phi_{0,B}^{(\Lambda)}\rangle\right|\nonumber\\
&\le&\frac{\epsilon}{\sqrt{\mathcal{E}_{k+Q}\mathcal{E}_k}}\nonumber\\
&+&\left(\frac{4S^2\mathcal{E}_{k+Q} + B S}{2\mathcal{E}_k}\right)^{1/4}
\sqrt{\langle\Phi_{0,B}^{(\Lambda)},\hat{S}_{-k}^{(2)}[\hat{g}(H_{0,B}^{(\Lambda)}-E_{0,B}^{(\Lambda)})]^2\hat{S}_{k}^{(2)}
\Phi_{0,B}^{(\Lambda)}\rangle+\frac{4S^2\mathcal{E}_k + B S}{\gamma-\Delta\gamma}}.\nonumber\\
\end{eqnarray}

As to the second term in the right-hand side of (\ref{msexpcommu}), one can see that 
the complex conjugate of the term satisfies 
$$
\overline{\langle\Phi_{0,B}^{(\Lambda)},\hat{S}_{Q+k}^{(3)}(1-P_0)\hat{S}_{-k}^{(2)}\Phi_{0,B}^{(\Lambda)}\rangle}
=\langle\Phi_{0,B}^{(\Lambda)},\hat{S}_{k}^{(2)}(1-P_0)\hat{S}_{Q-k}^{(3)}\Phi_{0,B}^{(\Lambda)}\rangle. 
$$
In addition, the reflection symmetry of the Hamiltonian $\tilde{H}_{0,B}^{(\Lambda)}$ of (\ref{tildeham}) 
implies that this right-hand side is equal to the quantity in the left-hand side of the above (\ref{msexpcommu1b}) 
which has the opposite sign for the momentum $k$. Consequently, we obtain 
\begin{eqnarray}
& &\frac{m_{B}^{(\Lambda)}}{2}-\frac{\epsilon}{\sqrt{\mathcal{E}_{k+Q}\mathcal{E}_k}}\nonumber\\
&\le& \left(\frac{4S^2\mathcal{E}_{k+Q} + B S}{2\mathcal{E}_k}\right)^{1/4}
\sqrt{\langle\Phi_{0,B}^{(\Lambda)},\hat{S}_{-k}^{(2)}[\hat{g}(H_{0,B}^{(\Lambda)}-E_{0,B}^{(\Lambda)})]^2\hat{S}_{k}^{(2)}
\Phi_{0,B}^{(\Lambda)}\rangle+\frac{4S^2\mathcal{E}_k + B S}{\gamma-\Delta\gamma}}.\nonumber\\
\end{eqnarray}
We choose the parameter $\epsilon=v_{\rm min}|p|$ with a small constant $v_{\rm min}>0$ so that 
\begin{equation}
\frac{m_{B}^{(\Lambda)}}{2}-\frac{\epsilon}{\sqrt{\mathcal{E}_{k+Q}\mathcal{E}_k}}>0
\end{equation}
for $k$ satisfying $|p|/2\le |k|\le |p|<\kappa$. Then, we have 
\begin{eqnarray}
\label{denomibound}
D^{(\Lambda)}(k,B)&:=&\left(\frac{2\mathcal{E}_k}{4S^2\mathcal{E}_{k+Q} + B S}\right)^{1/2}
\left[\frac{m_{B}^{(\Lambda)}}{2}-\frac{v_{\rm min}|p|}{\sqrt{\mathcal{E}_{k+Q}\mathcal{E}_k}}\right]^2
-\frac{4S^2\mathcal{E}_k + B S}{\gamma-\Delta\gamma}\nonumber\\
&\le& \langle\Phi_{0,B}^{(\Lambda)},\hat{S}_{-k}^{(2)}[\hat{g}(H_{0,B}^{(\Lambda)}-E_{0,B}^{(\Lambda)})]^2\hat{S}_{k}^{(2)}
\Phi_{0,B}^{(\Lambda)}\rangle
\end{eqnarray}
for $k$ satisfying $|p|/2\le |k|\le |p|<\kappa$.
Clearly, the left-hand side $D^{(\Lambda)}(k,B)$ can be strictly positive by choosing a small $v_{\rm min}>0$, a small $\kappa>0$ for  
the sufficiently weak field $B$. In addition, one has the lower bound, 
\begin{equation}
D(k,0):=\lim_{B\searrow 0}\lim_{\Lambda\nearrow\ze^d}D^{(\Lambda)}(k,B)\ge c_0|p|,
\end{equation}
with a constant $c_0>0$. 
Combining this, (\ref{supphatf}), (\ref{numerbound}), (\ref{denomi}) and (\ref{denomibound}), we obtain 
the lower and upper bounds for the excitation energy $\Delta E$ of (\ref{DeltaE}),  
\begin{eqnarray*}
v_{\rm min}|p|\le\Delta E&=&\lim_{B\searrow 0}\lim_{\Lambda\nearrow \ze^d}
\frac{\langle\Phi_{0,B}^{(\Lambda)},[\tau_{*g,B}^{(\Lambda)}(a_\Lambda)]^\ast[H_{0,B}^{(\Lambda)},\tau_{*g,B}^{(\Lambda)}(a_\Lambda)]
\Phi_{0,B}^{(\Lambda)}\rangle}{\langle\Phi_{0,B}^{(\Lambda)},[\tau_{*g,B}^{(\Lambda)}(a_\Lambda)]^\ast\tau_{*g,B}^{(\Lambda)}(a_\Lambda)]
\Phi_{0,B}^{(\Lambda)}\rangle}\\
&\le&\frac{(2\pi)^{-d}\int dk^{(1)}\int dk^{(2)}\cdots \int dk^{(d)}|\hat{f}(k)|^2\cdot 4S^2\mathcal{E}_k}{
(2\pi)^{-d}\int dk^{(1)}\int dk^{(2)}\cdots \int dk^{(d)}|\hat{f}(k)|^2D(k,0)}\\
&\le& \frac{2S^2|p|^2}{c_0|p|}=\frac{2S^2}{c_0}|p|,
\end{eqnarray*}
where we have used the quasi-locality of the observable $\lim_{\Lambda\nearrow \ze^d}\tau_{*g,B}^{(\Lambda)}(a)$, 
and the lower bound follows from (\ref{DeltaE}) with $\epsilon=v_{\rm min}|p|$.

%%%%%%%%%%%%%%%%%%%%%%%%%%%%%%%%%%%%%%%%%%%
\Section{A different Nambu-Goldstone mode}
\label{Sec:DifferentNGmode}

Instead of the observable $a_\Lambda$ of (\ref{aLambda}), we can choose 
\begin{equation}
\tilde{a}_\Lambda=\frac{1}{\sqrt{|\Lambda|}}\sum_k \hat{f}(k)\hat{S}_{k+Q}^{(2)}.
\end{equation}
This observable $\tilde{a}_\Lambda$ gives a low energy excitation with a linear dispersion relation around the momentum $Q$. 
When the corresponding observable $\mathcal{A}$ on the finite box $\Omega$ was already treated in \cite{KomaM},  
i.e., it is given by 
$$
\mathcal{A}=\frac{1}{|\Omega|}\sum_{x\in\Omega} (-1)^{x^{(1)}+\cdots+x^{(d)}}S_x^{(2)},
$$
and the existence of the Nambu-Goldstone mode was proved in \cite{KomaM}. 

In the same way as in (\ref{numerbound}), one has 
\begin{eqnarray*}
& &\langle \Phi_{0,B}^{(\Lambda)}, \tilde{a}_\Lambda^\ast[H_{0,B}^{(\Lambda)}-E_{0,B}^{(\Lambda)}]
[\hat{g}(H_{0,B}^{(\Lambda)}-E_{0,B}^{(\Lambda)})]^2\tilde{a}_\Lambda\Phi_{0,B}^{(\Lambda)}\rangle \\
&=&\frac{1}{|\Lambda|}\sum_k |\hat{f}(k)|^2\langle \Phi_{0,B}^{(\Lambda)}, \hat{S}_{-k+Q}^{(2)}
[H_{0,B}^{(\Lambda)}-E_{0,B}^{(\Lambda)}][\hat{g}(H_{0,B}^{(\Lambda)}-E_{0,B}^{(\Lambda)})]^2
\hat{S}_{k+Q}^{(2)}\Phi_{0,B}^{(\Lambda)}\rangle \\
&\le&\frac{1}{|\Lambda|}\sum_k |\hat{f}(k)|^2\langle \Phi_{0,B}^{(\Lambda)}, 
[\hat{S}_{-k+Q}^{(2)},[H_{0,B}^{(\Lambda)},\hat{S}_{k+Q}^{(2)}]]\Phi_{0,B}^{(\Lambda)}\rangle \\
&\le& \frac{1}{|\Lambda|}\sum_k |\hat{f}(k)|^2(4S^2\mathcal{E}_{k+Q}+BS).  
\end{eqnarray*}
The difference from the preceding case is that 
the quantity $\mathcal{E}_{k+Q}$ in the upper bound is non-vanishing in the limit $k\searrow 0$. 
Thus, roughly speaking, we want to show 
$$
\langle \Phi_{0,B}^{(\Lambda)}, \hat{S}_{-k+Q}^{(2)}
[\hat{g}(H_{0,B}^{(\Lambda)}-E_{0,B}^{(\Lambda)})]^2\hat{S}_{k+Q}^{(2)}\Phi_{0,B}^{(\Lambda)}\rangle\gtrsim \frac{1}{|k|}
$$
for a small $|k|$. 

For this purpose, let us consider the magnetization, 
\begin{equation}
m_B^{(\Lambda)}=-i\langle\Phi_{0,B}^{(\Lambda)},[\hat{S}_{-k+Q}^{(2)},\hat{S}_k^{(3)}]\Phi_{0,B}^{(\Lambda)}\rangle,
\end{equation}
similarly to the preceding case. Note that 
$$
\langle\Phi_{0,B}^{(\Lambda)},[\hat{S}_{-k+Q}^{(2)},\hat{S}_k^{(3)}]\Phi_{0,B}^{(\Lambda)}\rangle
=\langle\Phi_{0,B}^{(\Lambda)},\hat{S}_{-k+Q}^{(2)}(1-P_0)\hat{S}_k^{(3)}\Phi_{0,B}^{(\Lambda)}\rangle
-\langle\Phi_{0,B}^{(\Lambda)},\hat{S}_k^{(3)}(1-P_0)\hat{S}_{-k+Q}^{(2)}\Phi_{0,B}^{(\Lambda)}\rangle.
$$
Since both the first and second terms in the right-hand side can be treated in the same way as in the preceding case, 
we will consider only the first term in the following. It can be decomposed into two parts, 
\begin{eqnarray*}
& &\langle\Phi_{0,B}^{(\Lambda)},\hat{S}_{-k+Q}^{(2)}(1-P_0)\hat{S}_k^{(3)}\Phi_{0,B}^{(\Lambda)}\rangle\\
&=&\langle\Phi_{0,B}^{(\Lambda)},\hat{S}_{-k+Q}^{(2)}P(0,2\epsilon]\hat{S}_k^{(3)}\Phi_{0,B}^{(\Lambda)}\rangle
+\langle\Phi_{0,B}^{(\Lambda)},\hat{S}_{-k+Q}^{(2)}P(2\epsilon,\infty)\hat{S}_k^{(3)}\Phi_{0,B}^{(\Lambda)}\rangle.
\end{eqnarray*}
The first term in the right-hand side is evaluated as follows: 
$$
|\langle\Phi_{0,B}^{(\Lambda)},\hat{S}_{-k+Q}^{(2)}P(0,2\epsilon]\hat{S}_k^{(3)}\Phi_{0,B}^{(\Lambda)}\rangle|
\le \frac{\epsilon}{\sqrt{\mathcal{E}_{k+Q}\mathcal{E}_k}}
$$
in the same way as (\ref{contrib02epsilon}). As to the second term, one has 
\begin{eqnarray*}
|\langle\Phi_{0,B}^{(\Lambda)},\hat{S}_{-k+Q}^{(2)}P(2\epsilon,\infty)\hat{S}_k^{(3)}\Phi_{0,B}^{(\Lambda)}\rangle|
&\le&\sqrt{\langle\Phi_{0,B}^{(\Lambda)},\hat{S}_{-k+Q}^{(2)}P(2\epsilon,\infty)\hat{S}_{k+Q}^{(2)}\Phi_{0,B}^{(\Lambda)}\rangle}\\
&\times&\sqrt{\langle\Phi_{0,B}^{(\Lambda)},\hat{S}_{-k}^{(3)}P(2\epsilon,\infty)\hat{S}_k^{(3)}\Phi_{0,B}^{(\Lambda)}\rangle}.
\end{eqnarray*}
Further, similarly,  
\begin{eqnarray*}
\langle\Phi_{0,B}^{(\Lambda)},\hat{S}_{-k+Q}^{(2)}P(2\epsilon,\infty)\hat{S}_{k+Q}^{(2)}\Phi_{0,B}^{(\Lambda)}\rangle 
&\le&\langle\Phi_{0,B}^{(\Lambda)},\hat{S}_{-k+Q}^{(2)}[\hat{g}(H_{0,B}^{(\Lambda)}-E_{0,B}^{(\Lambda)})]^2\hat{S}_{k+Q}^{(2)}
\Phi_{0,B}^{(\Lambda)}\rangle\\
&+&\langle\Phi_{0,B}^{(\Lambda)},\hat{S}_{-k+Q}^{(2)}P[\gamma-\Delta\gamma,\infty)\hat{S}_{k+Q}^{(2)}\Phi_{0,B}^{(\Lambda)}\rangle\\
&\le&\langle\Phi_{0,B}^{(\Lambda)},\hat{S}_{-k+Q}^{(2)}[\hat{g}(H_{0,B}^{(\Lambda)}-E_{0,B}^{(\Lambda)})]^2\hat{S}_{k+Q}^{(2)}
\Phi_{0,B}^{(\Lambda)}\rangle\\
&+&\frac{4S^2\mathcal{E}_{k+Q}+BS}{\gamma-\Delta\gamma},
\end{eqnarray*}
and 
\begin{eqnarray*}
\langle\Phi_{0,B}^{(\Lambda)},\hat{S}_{-k}^{(3)}P(2\epsilon,\infty)\hat{S}_k^{(3)}\Phi_{0,B}^{(\Lambda)}\rangle
&\le&\sqrt{\langle\Phi_{0,B}^{(\Lambda)},\hat{S}_{-k}^{(3)}(H_{0,B}^{(\Lambda)}-E_{0,B}^{(\Lambda)})
\hat{S}_k^{(3)}\Phi_{0,B}^{(\Lambda)}\rangle}\\
&\times&\sqrt{\langle\Phi_{0,B}^{(\Lambda)},\hat{S}_{-k}^{(3)}\frac{1}{H_{0,B}^{(\Lambda)}-E_{0,B}^{(\Lambda)}}
\hat{S}_k^{(3)}\Phi_{0,B}^{(\Lambda)}\rangle}\\
&\le&\sqrt{\frac{4S^2\mathcal{E}_k+BS}{2\mathcal{E}_{k+Q}}}.
\end{eqnarray*}
Putting all together, we obtain 
\begin{eqnarray*}
& &\frac{m_B^{(\Lambda)}}{2}\le\frac{\epsilon}{\sqrt{\mathcal{E}_{k+Q}\mathcal{E}_k}}\\
&+&\left(\frac{4S^2\mathcal{E}_k+BS}{2\mathcal{E}_{k+Q}}\right)^{1/4}
\sqrt{\langle\Phi_{0,B}^{(\Lambda)},\hat{S}_{-k+Q}^{(2)}[\hat{g}(H_{0,B}^{(\Lambda)}-E_{0,B}^{(\Lambda)})]^2\hat{S}_{k+Q}^{(2)}
\Phi_{0,B}^{(\Lambda)}\rangle+\frac{4S^2\mathcal{E}_{k+Q}+BS}{\gamma-\Delta\gamma}}.
\end{eqnarray*}
We choose $\epsilon=v_{\rm min}|p|$ so that 
$$
\frac{m_B^{(\Lambda)}}{2}-\frac{\epsilon}{\sqrt{\mathcal{E}_{k+Q}\mathcal{E}_k}}>0
$$
for $k$ satisfying $|p|/2\le|k|\le|p|$ with a small $|p|$. Then, we have 
\begin{eqnarray*}
\langle\Phi_{0,B}^{(\Lambda)},\hat{S}_{-k+Q}^{(2)}[\hat{g}(H_{0,B}^{(\Lambda)}-E_{0,B}^{(\Lambda)})]^2\hat{S}_{k+Q}^{(2)}
\Phi_{0,B}^{(\Lambda)}\rangle&\ge& \left[\frac{m_B^{(\Lambda)}}{2}-\frac{\epsilon}{\sqrt{\mathcal{E}_{k+Q}\mathcal{E}_k}}\right]^2
\left(\frac{2\mathcal{E}_{k+Q}}{4S^2\mathcal{E}_k+BS}\right)^{1/2}\\
&-&\frac{4S^2\mathcal{E}_{k+Q}+BS}{\gamma-\Delta\gamma}.
\end{eqnarray*}
In the double limit, $\Lambda\nearrow\ze^d$ and $B\searrow 0$, the right-hand side shows the desired behavior $\sim 1/|k|$ 
for a small $|k|$.  

%%%%%%%%%%%%%%%%%%%%%%%%%%%%%%%%%%%%%%%%%%%%%%%
\appendix

\Section{Quasi-locality of the time evolution $\tau_{*g,B}$}
\label{Sec:QuasiLocal}

Following \cite{BMNS}, we decompose the time evolution $\tau_{*g,B}(\cdots)$ into a sum of local operators.

The present system satisfies the following Lieb-Robinson bound \cite{LR,NS,HK} for any operator $b$:   
\begin{equation}
\label{LRbound}
\left\Vert\left[\tau_{t,B}^{(\Lambda)}(a_x^{(n)}),b\right] \right\Vert \le K_1\Vert a_x^{(n)}\Vert \Vert b\Vert 
\sum_{y\in {\rm supp}\; b} e^{v|t|}e^{-\alpha|x-y|},
\end{equation}
where $K_1, v$ and $\alpha$ are some positive constants. We write $\Pi_X$ for 
the partial trace over the Hilbert space $\mathcal{H}_{X^c}$ on the complement $X^c$ of the sublattice $X$, i.e., 
\begin{equation}
\label{PiX}
\Pi_X(b):=\frac{1}{{\rm dim}\mathcal{H}_{X^c}}{\rm Tr}_{\mathcal{H}_{X^c}}b
\end{equation} 
for any operator $b$. Then, the Lieb-Robinson bound (\ref{LRbound}) yields \cite{BHV,BMNS,NSY}
\begin{equation}
\label{diffPitautau}
\left\Vert \Pi_X(\tau_{t,B}^{(\Lambda)}(a_x^{(n)}))-\tau_{t,B}^{(\Lambda)}(a_x^{(n)})\right\Vert 
\le 2K_1\Vert a_x^{(n)}\Vert \sum_{y\in X^c} e^{v|t|}e^{-\alpha|x-y|}. 
\end{equation}
Following \cite{BMNS}, we set 
\begin{equation}
\label{Xm(x)}
X_m(x):=\{y\; |\; |x-y|\le m\}\quad \mbox{for} \ \  m=0,1,2,\ldots,
\end{equation}
and 
\begin{equation}
\label{Delta0}
\Delta_0^{(\Lambda)}(a_x^{(n)}):=\int_{-\infty}^{+\infty}dt\; g(t)\cdot\Pi_{X_0(x)}(\tau_{t,B}^{(\Lambda)}(a_x^{(n)})),
\end{equation}
and 
\begin{equation}
\label{Deltam}
\Delta_m^{(\Lambda)}(a_x^{(n)}):=\int_{-\infty}^{+\infty}dt\; g(t)
\left[\Pi_{X_m(x)}(\tau_{t,B}^{(\Lambda)}(a_x^{(n)}))-\Pi_{X_{m-1}(x)}(\tau_{t,B}^{(\Lambda)}(a_x^{(n)}))\right]
\end{equation}
for $m\ge 1$. Then, one has 
\begin{equation}
\label{tauDeltasum}
\tau_{*g,B}^{(\Lambda)}(a_x^{(n)})=\int_{-\infty}^{+\infty}dt\; g(t)\cdot\tau_{t,B}^{(\Lambda)}(a_x^{(n)})
=\sum_{m=0}^\infty \Delta_m^{(\Lambda)}(a_x^{(n)}).  
\end{equation}
The summand satisfies  
\begin{equation}
\label{Deltabound}
\left\Vert\Delta_m^{(\Lambda)}(a_x^{(n)})\right\Vert\le K_2\Vert a_x^{(n)}\Vert \frac{1}{(m+1)^\beta},
\end{equation}
where $K_2$ and $\beta$ are positive constants.  
Since the positive exponent $\beta$ is determined by the decay of the function $g(t)$ for large $|t|$, 
we can take $\beta$ to be sufficiently large compared to the dimension $d$ of the lattice $\Lambda$. 
In addition, one has ${\rm supp}\; \Delta_m^{(\Lambda)}(a_x^{(n)})\subset X_m(x)$ by definition.  
Therefore, the quantity $\lim_{\Lambda\nearrow\ze^d}[H_{0,0}^{(\Lambda)},\tau_{*g,0}(a_x^{(n)})]$ exists 
and the norm is bounded because of the existence of $\tau_{*g,B}(\cdots)$ and of the continuity 
$\tau_{*g,0}(\cdots)=\lim_{B\searrow 0}\tau_{*g,B}(\cdots)$ (See Appendix~\ref{Sec:continuity}). 

Let us give the proof of the bound (\ref{Deltabound}). Note that 
\begin{eqnarray*}
& &\int_{-\infty}^{+\infty} dt\; g(t)\left[\Pi_{X_m(x)}(\tau_{t,B}^{(\Lambda)}(a_x^{(n)}))-\tau_{t,B}^{(\Lambda)}(a_x^{(n)})\right]\\
&=&\int_{|t|\le T} dt\; g(t)\left[\Pi_{X_m(x)}(\tau_{t,B}^{(\Lambda)}(a_x^{(n)}))-\tau_{t,B}^{(\Lambda)}(a_x^{(n)})\right]\\
&+&\int_{|t|>T} dt\; g(t)\left[\Pi_{X_m(x)}(\tau_{t,B}^{(\Lambda)}(a_x^{(n)}))-\tau_{t,B}^{(\Lambda)}(a_x^{(n)})\right] 
\end{eqnarray*}
for any $T>0$. By using the bound (\ref{diffPitautau}), the first term in the right-hand side can be estimated as follows: 
\begin{eqnarray*}
& &\left\Vert\int_{|t|\le T} dt\; g(t)\left[\Pi_{X_m(x)}(\tau_{t,B}^{(\Lambda)}(a_x^{(n)}))-\tau_{t,B}^{(\Lambda)}(a_x^{(n)})\right]
\right\Vert\\
&\le&\int_{|t|\le T} dt\; |g(t)|\left\Vert\Pi_{X_m(x)}(\tau_{t,B}^{(\Lambda)}(a_x^{(n)}))-\tau_{t,B}^{(\Lambda)}(a_x^{(n)}) \right\Vert\\
&\le& 2K_1\Vert a_x^{(n)}\Vert \int_{|t|\le T} dt\; |g(t)|\sum_{y\in X_m(x)^c} e^{vT}e^{-\alpha|x-y|}
\le K_3 \Vert a_x^{(n)}\Vert e^{vT}e^{-\alpha m}, 
\end{eqnarray*}
where $K_3$ is a positive constant. As to the second term, one has 
\begin{eqnarray*}
\left\Vert\int_{|t|>T} dt\; g(t)\left[\Pi_{X_m(x)}(\tau_{t,B}^{(\Lambda)}(a_x^{(n)}))-\tau_{t,B}^{(\Lambda)}(a_x^{(n)})\right]\right\Vert
&\le& \int_{|t|>T} dt\; |g(t)|\cdot 2\Vert a_x^{(n)})\Vert\\
&\le& K_4 \Vert a_x^{(n)})\Vert \frac{1}{T^\beta}, 
\end{eqnarray*}
where $K_4$ and $\beta$ are positive constants, and we have used that the function $g(t)$ decays by power law for a large $|t|$.  
By choosing $T=\alpha(m+1)/(2v)$, we have 
\begin{equation}
\label{Pitaudiff}
\left\Vert\int_{-\infty}^{+\infty} dt\; g(t)\left[\Pi_{X_m(x)}(\tau_{t,B}^{(\Lambda)}(a_x^{(n)}))-\tau_{t,B}^{(\Lambda)}(a_x^{(n)})
\right]\right\Vert\le {\rm Const.}\Vert a_x^{(n)})\Vert \frac{1}{(m+1)^\beta}.
\end{equation}
{From} this bound, one can obtain the desired bound (\ref{Deltabound}). 

%%%%%%%%%%%%%%%%%%%%%%%%%%%%%%%%%%%%%%%%%%%%%%%%%%%%%%%%%%
\Section{Continuity of the time evolution $\tau_{*g,B}$} 
\label{Sec:continuity}

In this appendix, we prove the inequality (\ref{contitauB}) below. 

Note that
\begin{eqnarray*}
\tau_{t,B}^{(\Lambda)}(a_x^{(n)})-\tau_{t,0}^{(\Lambda)}(a_x^{(n)})
&=&e^{iH_{0,B}^{(\Lambda)}t}a_x^{(n)}e^{-iH_{0,B}^{(\Lambda)}t}-e^{iH_{0,0}^{(\Lambda)}t}a_x^{(n)}e^{-iH_{0,0}^{(\Lambda)}t}\\
&=&\int_0^B du \frac{d}{du}e^{iH_{0,u}^{(\Lambda)}t}a_x^{(n)}e^{-iH_{0,u}^{(\Lambda)}t}\\ 
&=&\int_0^B du \left\{\left[\frac{d}{du}e^{iH_{0,u}^{(\Lambda)}t}\right]a_x^{(n)}e^{-iH_{0,u}^{(\Lambda)}t}
+e^{iH_{0,u}^{(\Lambda)}t}a_x^{(n)}\frac{d}{du}e^{-iH_{0,u}^{(\Lambda)}t}\right\}.
\end{eqnarray*}
The derivative in the right-hand side is calculated as follows: 
$$
\frac{d}{du}\; e^{iH_{0,u}^{(\Lambda)}t}=it \int_0^1 dw\; e^{iH_{0,u}^{(\Lambda)}tw}O^{(\Lambda)}e^{iH_{0,u}^{(\Lambda)}t(1-w)},
$$
where we have written 
$$
O^{(\Lambda)}:=\sum_{x\in\Lambda}(-1)^{x^{(1)}+x^{(2)}+\cdots+x^{(d)}}S_x^{(1)}.
$$
Using this formula, the above integrand is computed as 
\begin{eqnarray*}
& &\left[\frac{d}{du}e^{iH_{0,u}^{(\Lambda)}t}\right]a_x^{(n)}e^{-iH_{0,u}^{(\Lambda)}t}
+e^{iH_{0,u}^{(\Lambda)}t}a_x^{(n)}\frac{d}{du}e^{-iH_{0,u}^{(\Lambda)}t}\\
&=&it \int_0^1 dw\left[e^{iH_{0,u}^{(\Lambda)}tw}O^{(\Lambda)}e^{iH_{0,u}^{(\Lambda)}t(1-w)}a_x^{(n)}e^{-iH_{0,u}^{(\Lambda)}t}
-e^{iH_{0,u}^{(\Lambda)}t}a_x^{(n)}e^{-iH_{0,u}^{(\Lambda)}t(1-w)}O^{(\Lambda)}e^{-iH_{0,u}^{(\Lambda)}tw}\right]\\
&=&it \int_0^1 dw\; e^{iH_{0,u}^{(\Lambda)}tw}\left[O^{(\Lambda)},e^{iH_{0,u}^{(\Lambda)}t(1-w)}a_x^{(n)}e^{-iH_{0,u}^{(\Lambda)}t(1-w)}
\right]e^{-iH_{0,u}^{(\Lambda)}tw}\\
&=&it \int_0^1 dw\; e^{iH_{0,u}^{(\Lambda)}tw}\left[O^{(\Lambda)},\tau_{t(1-w),u}^{(\Lambda)}(a_x^{(n)})\right]e^{-iH_{0,u}^{(\Lambda)}tw}
\end{eqnarray*}
{From} these observations, we obtain 
\begin{eqnarray}
\label{taugBdiff}
& &\tau_{*g,B}^{(\Lambda)}(a_x^{(n)})-\tau_{*g,0}^{(\Lambda)}(a_x^{(n)})\nonumber\\
&=&\int_{-\infty}^{+\infty}dt\; g(t)\left[\tau_{t,B}^{(\Lambda)}(a_x^{(n)})-\tau_{t,0}^{(\Lambda)}(a_x^{(n)})\right]\nonumber\\
&=&i\int_{-\infty}^{+\infty}dt\; tg(t)\int_0^B du \int_0^1 dw\; e^{iH_{0,u}^{(\Lambda)}tw}
\left[O^{(\Lambda)},\tau_{t(1-w),u}^{(\Lambda)}(a_x^{(n)})\right]e^{-iH_{0,u}^{(\Lambda)}tw}.
\end{eqnarray}
In order to estimate the right-hand side, we set 
$$
\tilde{\Delta}_0^{(\Lambda)}(\tau_{t(1-w),u}^{(\Lambda)}(a_x^{(n)})):=\Pi_{X_0(x)}(\tau_{t(1-w),u}^{(\Lambda)}(a_x^{(n)})),
$$
and 
$$
\tilde{\Delta}_m^{(\Lambda)}(\tau_{t(1-w),u}^{(\Lambda)}(a_x^{(n)}))
:=\Pi_{X_m(x)}(\tau_{t(1-w),u}^{(\Lambda)}(a_x^{(n)}))-\Pi_{X_{m-1}(x)}(\tau_{t(1-w),u}^{(\Lambda)}(a_x^{(n)}))
$$
for $m\ge 1$. Then, one has 
$$
\tau_{t(1-w),u}^{(\Lambda)}(a_x^{(n)})=\sum_{m=0}^\infty \tilde{\Delta}_m^{(\Lambda)}(\tau_{t(1-w),u}^{(\Lambda)}(a_x^{(n)})).
$$
Substituting this into the right-hand side of above (\ref{taugBdiff}), we have 
\begin{eqnarray}
\label{tauBdiffSum}
& &i\int_{-\infty}^{+\infty}dt\; tg(t)\int_0^B du\; \int_0^1 dw\; e^{iH_{0,u}^{(\Lambda)}tw}
\left[O^{(\Lambda)},\tau_{t(1-w),u}^{(\Lambda)}(a_x^{(n)})\right]e^{-iH_{0,u}^{(\Lambda)}tw}\nonumber\\
&=&i \sum_{m=0}^\infty \int_{-\infty}^{+\infty}dt\; tg(t)\int_0^B du\; \int_0^1 dw\; e^{iH_{0,u}^{(\Lambda)}tw}
\left[O^{(\Lambda)},\tilde{\Delta}_m^{(\Lambda)}(\tau_{t(1-w),u}^{(\Lambda)}(a_x^{(n)}))\right]e^{-iH_{0,u}^{(\Lambda)}tw}.\nonumber\\
\end{eqnarray}
This summand for $m\ge 1$ is written 
\begin{eqnarray}
\label{tauBdiffSummand}
& &\int_{-\infty}^{+\infty}dt\; tg(t)\int_0^B du\; \int_0^1 dw\; e^{iH_{0,u}^{(\Lambda)}tw}
\left[O^{(\Lambda)},\tilde{\Delta}_m^{(\Lambda)}(\tau_{t(1-w),u}^{(\Lambda)}(a_x^{(n)}))\right]e^{-iH_{0,u}^{(\Lambda)}tw}\nonumber\\
&=&\int_{|t|\le T}dt\; tg(t)\int_0^B du\; \int_0^1 dw\; e^{iH_{0,u}^{(\Lambda)}tw}
\left[O^{(\Lambda)},\tilde{\Delta}_m^{(\Lambda)}(\tau_{t(1-w),u}^{(\Lambda)}(a_x^{(n)}))\right]e^{-iH_{0,u}^{(\Lambda)}tw}\nonumber\\
&+&\int_{|t|>T}dt\; tg(t)\int_0^B du\; \int_0^1 dw\; e^{iH_{0,u}^{(\Lambda)}tw}
\left[O^{(\Lambda)},\tilde{\Delta}_m^{(\Lambda)}(\tau_{t(1-w),u}^{(\Lambda)}(a_x^{(n)}))\right]e^{-iH_{0,u}^{(\Lambda)}tw}. 
\end{eqnarray}
Since the support of the operator $\tilde{\Delta}_m^{(\Lambda)}(\tau_{t(1-w),u}^{(\Lambda)}(a_x^{(n)}))$ is contained in 
the set $X_m(x)$ by definition, we have 
\begin{equation}
\label{Ocommubound}
\left\Vert\left[O^{(\Lambda)},\tilde{\Delta}_m^{(\Lambda)}(\tau_{t(1-w),u}^{(\Lambda)}(a_x^{(n)}))\right]\right\Vert
\le \tilde{K}_1 m^d \left\Vert\tilde{\Delta}_m^{(\Lambda)}(\tau_{t(1-w),u}^{(\Lambda)}(a_x^{(n)}))\right\Vert, 
\end{equation}  
where $\tilde{K}_1$ is a positive constant which depends on the dimension $d$ of the lattice $\Lambda$. 

Consider first the first term in the right-hand side of (\ref{tauBdiffSummand}).
Similarly to the preceding Appendix~\ref{Sec:QuasiLocal}, one has 
\begin{equation}
\left\Vert\tilde{\Delta}_m^{(\Lambda)}(\tau_{t(1-w),u}^{(\Lambda)}(a_x^{(n)}))\right\Vert
\le \tilde{K}_2 \Vert a_x^{(n)}\Vert e^{vT}e^{-\alpha m}
\end{equation}
with a positive constant $\tilde{K}_2$ for $|t|\le T$. Combining this with the above bound (\ref{Ocommubound}), 
the first term in the right-hand side of (\ref{tauBdiffSummand}) can be estimated as follows: 
\begin{eqnarray}
& &\left\Vert\int_{|t|\le T}dt\; tg(t)\int_0^B du\; \int_0^1 dw\; e^{iH_{0,u}^{(\Lambda)}tw}
\left[O^{(\Lambda)},\tilde{\Delta}_m^{(\Lambda)}(\tau_{t(1-w),u}^{(\Lambda)}(a_x^{(n)}))\right]e^{-iH_{0,u}^{(\Lambda)}tw}
\right\Vert\nonumber\\
&\le&\tilde{K}_3|B|\Vert a_x^{(n)}\Vert m^d e^{vT}e^{-\alpha m}
\end{eqnarray}
with a positive constant $\tilde{K}_3$. 

For the second term in the right-hand side of (\ref{tauBdiffSummand}), we use the inequality, 
$$
\int_{|t|>T}dt\; |t||g(t)|\le {\rm Const.}\frac{1}{T^{\tilde{\beta}}},  
$$
with a positive constant $\tilde{\beta}$. This holds, and we can take a sufficiently large $\tilde{\beta}$ 
because of the assumption of the infinite differentiability of the function $\hat{g}$. 
Combining this inequality with the above bound (\ref{Ocommubound}), one has 
\begin{eqnarray*}
& &\left\Vert\int_{|t|>T}dt\; tg(t)\int_0^B du\; \int_0^1 dw\; e^{iH_{0,u}^{(\Lambda)}tw}
\left[O^{(\Lambda)},\tilde{\Delta}_m^{(\Lambda)}(\tau_{t(1-w),u}^{(\Lambda)}(a_x^{(n)}))\right]e^{-iH_{0,u}^{(\Lambda)}tw}\right\Vert\\ 
&\le& \tilde{K}_4\frac{1}{T^{\tilde{\beta}}}|B|
m^d \left\Vert\tilde{\Delta}_m^{(\Lambda)}(\tau_{t(1-w),u}^{(\Lambda)}(a_x^{(n)}))\right\Vert
\le 2\tilde{K}_4|B|\Vert a_x^{(n)}\Vert\frac{m^d}{T^{\tilde{\beta}}}  
\end{eqnarray*}
with a positive constant $\tilde{K}_4$. By choosing $T=\alpha(m+1)/(2v)$, we consequently obtain 
\begin{eqnarray*}
& &\left\Vert\int_{-\infty}^{+\infty}dt\; tg(t)\int_0^B du\; \int_0^1 dw\; e^{iH_{0,u}^{(\Lambda)}tw}
\left[O^{(\Lambda)},\tilde{\Delta}_m^{(\Lambda)}(\tau_{t(1-w),u}^{(\Lambda)}(a_x^{(n)}))\right]e^{-iH_{0,u}^{(\Lambda)}tw}\right\Vert\\
&\le& \tilde{K}_5|B| \Vert a_x^{(n)}\Vert \frac{1}{(m+1)^{\tilde{\beta}-d}}
\end{eqnarray*}
with a positive constant $\tilde{K}_5$. Substituting this into (\ref{tauBdiffSum}), we have 
\begin{eqnarray*}
& &\left\Vert\int_{-\infty}^{+\infty}dt\; tg(t)\int_0^B du\; \int_0^1 dw\; e^{iH_{0,u}^{(\Lambda)}tw}
\left[O^{(\Lambda)},\tau_{t(1-w),u}^{(\Lambda)}(a_x^{(n)})\right]e^{-iH_{0,u}^{(\Lambda)}tw}\right\Vert\\
&\le&\tilde{K}_5|B| \Vert a_x^{(n)}\Vert \sum_{m=0}^\infty \frac{1}{(m+1)^{\tilde{\beta}-d}}
\le \tilde{K}_6 |B| \Vert a_x^{(n)}\Vert
\end{eqnarray*}
with a positive constant $\tilde{K}_6$, where we have chosen $\tilde{\beta}$ so that 
$\tilde{\beta}-d>1$. Therefore, from (\ref{taugBdiff}), we obtain 
\begin{equation}
\left\Vert\tau_{*g,B}^{(\Lambda)}(a_x^{(n)})-\tau_{*g,0}^{(\Lambda)}(a_x^{(n)})\right\Vert
\le \tilde{K}_6 |B| \Vert a_x^{(n)}\Vert. 
\end{equation}
This is the desired continuity at $B=0$ with respect to $B$. Since the constant $\tilde{K}_6$ does not depend on 
the lattice $\Lambda$, one has 
\begin{equation}
\label{contitauB}
\left\Vert\tau_{*g,B}(a_x^{(n)})-\tau_{*g,0}(a_x^{(n)})\right\Vert
\le \tilde{K}_6 |B| \Vert a_x^{(n)}\Vert. 
\end{equation}
in the infinite-volume limit.

%%%%%%%%%%%%%%%%%%%%%%%%%%%%%%%%%%%%%%%%%%%%%%%%%%%%%%%%%%%%%%%%%%%%%%%%%%%
\bigskip\bigskip\bigskip

\noindent
{\bf Acknowledgements:} I would like to thank Akinori Tanaka for sending his note and for helpful comments. 
%%%%%%%%%%%%%%%%%%%%%%%%%%%%%%%%%%%%%%%%%%%%%%%%%%%%%%%%%%%%%
%\newpage

\end{document}